\documentclass{article}

\usepackage{arxiv}                    
\usepackage{amsmath,amssymb}          
\usepackage{amsfonts}                 
\usepackage[utf8]{inputenc}           
\usepackage[T1]{fontenc}              

\usepackage{microtype}                
\usepackage{url}                      
\usepackage{xcolor}                   

\usepackage{booktabs}                 
\usepackage{graphicx}                 
\usepackage{nicefrac}                 

\usepackage{natbib}                   
\usepackage{doi}                      
\usepackage{cleveref}                 

\usepackage{hyperref}                 
\hypersetup{
	hidelinks,                        
	colorlinks = true,
	citecolor = cyan,
	urlcolor = black,
	linkcolor = magenta
}

\usepackage{dsfont}
\usepackage{subcaption}

\usepackage{algorithm}                
\usepackage{algpseudocode}            

\newcommand{\E}{\mathbb{E}}           

\newcommand{\abs}[1]{\left|#1\right|}            

\title{Multiscale Modelling of Birth-Death Processes}

\newif\ifuniqueAffiliation

\ifuniqueAffiliation
\author{
	\href{https://orcid.org/0000-0000-0000-0000}{\includegraphics[scale=0.06]{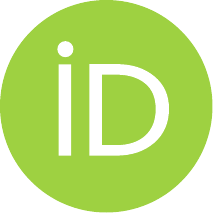}\hspace{1mm}Author 1}\thanks{Use footnote for providing further information about author (webpage, alternative address)---\emph{not} for acknowledging funding agencies.} \\
	Department of X\\
	Y University\\
	somewhere in the world \\
	\texttt{me@university.edu} \\
	\And
	\href{https://orcid.org/0000-0000-0000-0000}{\includegraphics[scale=0.06]{orcid.pdf}\hspace{1mm}Elias D.~Striatum} \\
	Department of Electrical Engineering\\
	Mount-Sheikh University\\
	Santa Narimana, Levand \\
	\texttt{stariate@ee.mount-sheikh.edu}
}
\else
\usepackage{authblk}

\newbox{\orcid}\sbox{\orcid}{\includegraphics[scale=0.06]{orcid.pdf}}

\author[1,2]{%
	\href{https://orcid.org/0000-0002-6542-6032}{\usebox{\orcid}\hspace{1mm}Tom Kimpson\thanks{Corresponding author: \texttt{tom.kimpson@unimelb.edu.au}}}%
}

\author[1]{%
	\href{https://orcid.org/0000-0001-5893-4840}{\usebox{\orcid}\hspace{1mm}Domenic P.J. Germano}%
}

\author[1,2]{%
	\href{https://orcid.org/0000-0002-8809-726X}{\usebox{\orcid}\hspace{1mm}Jennifer A. Flegg}%
}

\author[3]{%
	\href{https://orcid.org/0000-0002-4697-4789}{\usebox{\orcid}\hspace{1mm}Mark B. Flegg}%
}

\affil[1]{School of Mathematics and Statistics, The University of Melbourne, Parkville, VIC 3010, Australia}
\affil[2]{ARC Centre of Excellence for the Mathematical Analysis of Cellular Systems (MACSYS), The University of Melbourne, Parkville, VIC 3010, Australia }

\affil[3]{School of Mathematics, Monash University, Clayton, VIC 3800, Australia}

\fi

\renewcommand{\headeright}{}          
\renewcommand{\undertitle}{}          
\renewcommand{\shorttitle}{Birth-death multiscale modelling}

\hypersetup{
	pdftitle={Multiscale Modelling of Birth-Death Processes},
	pdfsubject={q-bio.NC, q-bio.QM},
	pdfauthor={Tom Kimpson},
	pdfkeywords={Birth-death processes, Multiscale modelling, Stochastic processes}
}

\begin{document}
\maketitle

	\begin{abstract}
	 Many biological systems exhibit multiscale dynamics, where some species occur in high copy numbers while others remain rare. This heterogeneity necessitates hybrid modelling approaches: deterministic models are computationally efficient but inaccurate for low-count species, while fully stochastic simulations are accurate but prohibitively expensive. Threshold-based hybrid methods, such as the Jump--Switch--Flow (JSF) algorithm, address this by simulating low-count species stochastically and high-count species deterministically, switching at a user-chosen threshold $\Omega$. In such methods, the choice of $\Omega$ controls the trade-off between computational cost and accuracy, but is typically made by trial-and-error: there is no principled way to choose $\Omega$ \emph{a priori} for a given observable of interest. We close this gap for extinction probability. Our contribution is a computable, method-agnostic error bound that quantifies the discrepancy introduced by the threshold and yields an explicit rule for selecting $\Omega$ to meet a user-specified error tolerance. We formalise JSF as a piecewise-deterministic Markov process and derive backward equations for extinction under exact and hybrid dynamics. Near extinction boundaries, the complex nonlinear dynamics reduce to tractable time-inhomogeneous linear birth--death processes; this structure yields a rigorous error decomposition into early and late excursions, whose dominant term becomes a fast, actionable heuristic requiring only the solution of a scalar Riccati equation. Monte Carlo studies on a stochastic Lotka--Volterra model confirm that the heuristic reliably upper-bounds the empirical error in extinction probability across a wide parameter range. The framework depends only on the birth and death rates near extinction, not on the specific simulation method, and therefore applies beyond JSF to any threshold-based hybrid scheme.
	\end{abstract}


\section{Introduction}
Population dynamics in biological systems frequently navigate vast orders of magnitude across temporal and spatial scales. In epidemiology, infectious diseases emerge from molecular interactions within individual hosts, spread through local transmission networks, and can ultimately manifest as global pandemics affecting millions \citep{Childs2019,Doran2023,Wang2022}. In ecology, populations demonstrate dramatic ``boom-and-bust'' cycles where abundances fluctuate across orders of magnitude \citep{Ludwig1978,Levin1992,Nichols2008,McGarigal_2016}. In systems biology, cellular and molecular processes cascade through tissue-level organization to produce whole-organism phenotypes \citep{Dada2011,Walpole2013,Qu2011}.

At the finest scale, discrete-state, continuous-time Markov chains (CTMCs) provide the most physically accurate description of spatially well mixed biological systems, explicitly modelling the probabilistic nature of individual events. The foundational method for simulating CTMCs is the Doob-Gillespie Stochastic Simulation Algorithm (SSA) \citep{Doob1942TopicsMarkoffChains,Doob1945MarkoffChainsDenumerable,Gillespie1976GeneralMethodSSA,Gillespie1977ExactSSA}, which serves as the ``gold standard'' for accuracy by capturing the true stochastic dynamics of discrete biological processes. However, the SSA's computational cost presents a significant barrier to practical application. The algorithm simulates every single event, making it prohibitively computationally expensive for systems involving large populations or rapid reaction rates, rendering it impractical for most realistic biological systems.

To address these computational limitations while retaining stochasticity, several fast-but-approximate methods have been developed, such as the tau-leaping algorithm \citep{Gillespie2001TauLeaping} and the chemical Langevin equation \citep{Gillespie2000CLE,Rao2003CLE,Cao2006CLE,Gibson2000CLE,Simoni2019SSAReview}. These approaches accelerate simulation by approximating multiple reaction events collectively or by treating fluctuations as continuous noise. However, while they can be effective in certain regimes, the error incurred by these approximations is often unacceptable when modelling systems where rare events, absorption states, or precise stochastic dynamics play a critical role.

At larger scales, continuous deterministic models such as ordinary differential equations (ODEs) provide an efficient alternative. ODEs are computationally efficient and effective for systems with large numbers of interacting components (such as vast populations of cells, molecules, or organisms) because at such large scales, random fluctuations in individual events (births, deaths, reactions) tend to average out, justifying the continuum approach. Indeed, many ODEs used in biology can be mathematically justified as the ``ensemble average'' behaviour of an underlying stochastic process in the limit of large populations \citep{kurtz1970solutions,kurtz1971limit,kurtz1972relationship}. However, this continuum assumption comes with significant drawbacks. At smaller scales, the inherent stochasticity of individual events can no longer be averaged out, and the continuous ODE treatment of discrete processes can lead to artefacts such as ``atto-fox problems'' where populations shrink to unphysically small sizes that persist indefinitely, whereas in reality the last individual would have died out \citep{Fowler2021AttoFoxes,LobrySari2015}. Accordingly, ODE models are incapable of meaningfully addressing threshold-sensitive or stochasticity-driven phenomena. A prime example is absorption into extinction states which are ubiquitous in biological systems yet fundamentally require stochastic treatment.

When the system dynamics remains firmly in a single large-scale or small-scale regime, the choice between continuous or discrete modelling approaches is clear. However, for multiscale systems that traverse both large and small population regimes, neither approach is sufficient on its own. This fundamental gap creates the central need for hybrid methodologies, as deterministic models are inaccurate for small populations and stochastic models become computationally infeasible for large ones. Developing methods that can faithfully bridge these scales has been a long-standing goal in applied mathematics \citep{cotter2016error,flegg2014analysis,isaacson2013convergent}.

To bridge the gap between deterministic and stochastic approaches, a variety of hybrid modelling methods have been developed. These methods aim to provide a ``best of both worlds'' solution by dynamically partitioning a system and applying the most appropriate simulation technique to each partition, thereby combining the accuracy of stochastic models for small populations with the speed of deterministic models for large ones. Classic approaches include partitioned hybrid schemes which approximates fast reactions deterministically while simulating slow reactions stochastically \citep{Haseltine2002}, and partitioned leaping methods that categorise reactions as fast or slow \citep{Cao2005Partition}. A parallel line of work partitions by \emph{species abundance} using population thresholds, typically augmented with a middle Chemical Langevin Equation (CLE) regime between the SSA and the deterministic ODE, as in the multi-level schemes of \citet{WinkelmannSchutte2017,WinkelmannSchutte2020}; see \citet{WinkelmannSchutte2020} for a comprehensive review. Further innovations include the regime-conversion method \citep{Kynaston2023}; jump--diffusion processes \citep{Buckwar2011, ANGIUS201549}; piecewise deterministic Markov processes (PDMPs) \citep{Alfonsi2005Adaptive, Marchetti2016HRSSA, Crudu2009Hybrid}; and the ``Jump-Switch-Flow'' method \citep{2024arXiv240513239G}. For a broader review of hybrid multiscale modelling approaches, we refer the interested reader to \citet{Simoni2019SSAReview}.

One practical question recurs across these methods: how should the threshold $\Omega$ be chosen \emph{a priori} to control the error in a chosen observable? In practice $\Omega$ is fixed by trial-and-error, or by heuristics that cannot be checked against a target tolerance. The contribution of this paper is not a new partitioning scheme. It is a computable error bound for extinction probability that answers this question, and that holds regardless of the simulator. We use JSF as the simulator throughout, but the analysis depends only on the birth and death rates of the scarce species near the extinction boundary.

To address the computational challenges of multiscale biological systems, in this manuscript we employ the Jump-Switch-Flow (JSF) hybrid method. We choose the JSF framework for three reasons. Firstly, it offers a dramatic speed-up compared to a full CTMC simulation, making it feasible to perform computationally intensive tasks like Bayesian parameter inference, which require thousands of model simulations. Secondly, its ability to accurately model extinction events by retaining a discrete, integer-based population count is critical for the biological questions we address. Unlike ODEs, it avoids the ``atto-fox'' problem, and unlike some tau-leaping methods, it will not produce negative populations when properly calibrated. Finally, its implementation is more straightforward than many other hybrid methods, as it avoids the need to partition reaction channels or solve complex stochastic differential equations, and is readily available as a user-friendly public package \footnote{\url{https://dgermano8.github.io/JSF/}}. The JSF method provides a pragmatic and powerful balance of biological realism and computational tractability for multiscale population systems. Broadly, JSF has two regimes, separated by a user-chosen vector of integer switching thresholds $\boldsymbol{\Omega}=(\Omega_1,\dots,\Omega_N)\in\mathbb{N}^N$, with one entry $\Omega_i$ per species:
	\begin{itemize}
		\item \textbf{Stochastic Regime (Jump):} When the number of individuals in compartment $i$ ($N_i$) is low ($N_i \le \Omega_i$), that compartment's dynamics are governed by a stochastic CTMC process. Events (e.g., birth, death, infection) are simulated exactly in the spirit of the SSA \citep{Gillespie1977ExactSSA}, capturing the intrinsic noise that is critical at low population levels. This ensures that true extinction (the population reaching zero) is possible and accurately modelled.
		\item \textbf{Deterministic Regime (Flow):} When the number of individuals in a compartment $i$ is high ($N_i > \Omega_i$), that compartment's dynamics are modelled using a deterministic ODE. This ``flow'' approximation is computationally fast and accurately reflects the average behaviour of the large population.
	\end{itemize}	  	
 A central unresolved question in applying the JSF method is how to choose the threshold parameter $\Omega_i$. In \citet{2024arXiv240513239G}, $\Omega_i$ was set heuristically by running trial simulations and selecting a value that resulted in accurate, but reasonable computational performance. In this work, we take a different approach: we aim to provide a systematic approach for selecting $\Omega_i$ \textit{a priori}, without the need for extensive exploratory runs. 
 
We note that the ``optimal'' choice of $\Omega_i$ is not universal: it depends on the biological or epidemiological question under investigation, as well as the willingness to sacrifice accuracy for computational efficiency For example, if the research question concerns accurately quantifying the probabilities of extinction or the timing of rare stochastic events, a relatively low $\Omega_i$ is often sufficient. This choice maximises computational efficiency by using the deterministic model at high counts, while critically retaining the essential stochastic dynamics as the population approaches the extinction boundaries. Conversely, if the primary aim is to capture complex dynamics like stochastic cycles, or if the mean-field model is known to be a poor approximation even at high population sizes, a larger $\Omega_i$ may be necessary. This ensures that stochastic dynamics are retained over a much broader range of the state space, at the cost of computational speed.

In this study, we focus exclusively on the objective of quantifying extinction probabilities accurately within a finite time horizon. We develop a systematic method for selecting a regime switching threshold, $\Omega_i$, that balances computational efficiency and accuracy: we seek the smallest threshold that keeps the error between the exact SSA solution and the JSF hybrid solution within a prescribed tolerance. Specifically, we derive a method to calculate an upper bound on the error in extinction probability introduced by the hybrid approximation. We focus on extinction events due to their ubiquitous biological and epidemiological significance. Extinction events, including viral clearance, species extinction, and the eradication of invasive populations, are fundamentally predicated on low-population stochasticity. This makes extinction an ideal case study to demonstrate how a principled selection of $\Omega_i$ can be tailored to a specific scientific objective.

Our analytical approach exploits a key simplification: near extinction boundaries, the complex nonlinear dynamics of realistic biological systems reduce to tractable birth-death processes. When populations are small, interactions between rare individuals become negligible, and each individual's fate can be analysed independently. This birth-death structure enables exact computation of extinction probabilities via classical branching process theory, and provides the mathematical foundation for our error analysis.


This paper is organised as follows. Section~\ref{sec:general_framework} introduces the three modelling frameworks: deterministic ODEs, stochastic CTMCs as the gold standard of accuracy, and JSF hybrid methods. Section~\ref{sec:birth-death-framework} develops the mathematical framework underpinning our analysis, showing how general systems reduce to birth-death processes near extinction; this reduction is the key that enables the error bounds derived in Section~\ref{sec:error_analysis}. Although we work with JSF throughout for concreteness, the derivation depends only on the birth and death rates of the scarce species near the extinction boundary, so the resulting bounds apply to any threshold-based hybrid scheme; we return to this point in Section~\ref{sec:discussion}. Building on this foundation, Section~\ref{sec:error_analysis} derives theoretical error bounds and practical heuristics for hybrid threshold selection. Section~\ref{sec:lv-case-study} demonstrates the theory through a detailed case study using the stochastic Lotka-Volterra predator-prey system, with explicit solutions and numerical validation. Lastly, we discuss the broader implications of the work and future extensions in Section~\ref{sec:discussion}.

\section{Modelling frameworks}\label{sec:general_framework}

This section outlines the three mathematical frameworks used to model population dynamics: the exact stochastic description, the deterministic large-population limit, and a hybrid method that combines elements of both.

\subsection{The Stochastic Model: Continuous-Time Markov Chains (CTMC)}

The most faithful representation of stochastic events occurring in a well-mixed population system is by a Continuous-Time Markov Chain (CTMC). Let the state of the system at time $t \ge 0$ be the vector of species populations, $\mathbf{X}(t)=(X_1(t),\dots,X_N(t)) \in \mathbb{N}_0^N$. The system evolves through a set of $R$ possible reactions. Each reaction $r$ is defined by:
\begin{itemize}
	\item A \textbf{stoichiometric vector} $\boldsymbol{\nu}_r \in \mathbb{Z}^N$, which describes the change in population counts when reaction $r$ occurs.
	\item A \textbf{propensity function} $a_r(\mathbf{x},t) \ge 0$, which gives the instantaneous probability rate of reaction $r$ occurring, given the system is in state $\mathbf{x}$ at time $t$.
\end{itemize}
The dynamics of the CTMC are fully captured by its generator, $\mathcal{L}_t$. Acting on a bounded test function $f:\mathbb{N}_0^N \to \mathbb{R}$, the generator describes the instantaneous expected rate of change of $f(\mathbf{X}(t))$ \citep{andersonkurtz}
\begin{equation}\label{eq:ctmc_generator}
	(\mathcal{L}_t f)(\mathbf{x}) = \sum_{r=1}^R a_r(\mathbf{x},t) \big[f(\mathbf{x}+\boldsymbol{\nu}_r)-f(\mathbf{x})\big].
\end{equation}

The generator is related, by duality, to the forward (Kolmogorov) operator ($\mathcal{L}_t^\ast$), which acts on the probability distribution $(p(\mathbf{x},t) = \Pr[\mathbf{X}(t)=\mathbf{x}])$
\begin{equation}
\frac{d}{dt} p(\mathbf{x},t) = (\mathcal{L}_t^\ast p)(\mathbf{x},t)= \sum_{r=1}^R \big[a_r(\mathbf{x}-\boldsymbol{\nu}_r,t)p(\mathbf{x}-\boldsymbol{\nu}_r,t) - a_r(\mathbf{x},t)p(\mathbf{x},t)\big],
\end{equation}
with the generator and its adjoint satisfying
\begin{equation}
 \frac{d}{dt}\mathbb{E}[f(\mathbf{X}(t))] = \mathbb{E}[(\mathcal{L}_t f)(\mathbf{X}(t))].
\end{equation}
Events such as extinction are handled by defining an absorbing set $\mathcal{A} \subseteq \mathbb{N}_0^N$, where the process stops once it enters. While this approach provides our most accurate description of stochastic population dynamics, simulating every individual event becomes computationally prohibitive for large populations.

\subsection{The Deterministic Limit: Mean-Field ODEs}
In systems where all population counts are very large, stochastic effects become negligible. In this regime, the law of large numbers guarantees that the system's behaviour converges to a deterministic path. This path is described by a set of Ordinary Differential Equations (ODEs), often called the mean-field or Kurtz limit:
\begin{equation}\label{eq:ode_limit}
	\dot{\mathbf{x}}(t) = \sum_{r=1}^R \boldsymbol{\nu}_r \, a_r\big(\mathbf{x}(t),t\big), \qquad \text{with initial condition} \quad \mathbf{x}(0)=\mathbf{x}_0 \in \mathbb{R}_{\ge 0}^N.
\end{equation}
While computationally efficient, with standard numerical methods such as Euler and Runge-Kutta readily available, this model completely ignores the random fluctuations that are critical for describing phenomena like extinction in low-population systems. Despite the name, mean-field ODE models do not necessarily describe the mean behaviour of the CTMC; in fact, for non-linear systems, they rarely do.

\subsection{The Hybrid Model: Jump-Switch-Flow (JSF)}\label{subsec:JSF}
Neither the CTMC nor ODE approach alone are sufficient for describing multiscale biological systems. The core challenge is that these systems often exhibit a mixture of reactions occurring on vastly different timescales (i.e., `fast' and `slow' reactions simultaneously). This `stiffness' problem is then further compounded by populations that can transition between high-count (continuum) and low-count (discrete) regimes. The JSF process \citep{2024arXiv240513239G} addresses these challenges by dynamically blending the accuracy of the CTMC at low populations with the efficiency of the ODE model at large populations. The core idea is to partition the $N$ species compartments dynamically based on a vector of integer thresholds, $\boldsymbol\Omega=(\Omega_1,\dots,\Omega_N) \in \mathbb{N}^N$. At any time $t$, we define two index sets:
\begin{equation}\label{eq:jsf_sets}
	J(t)=\{i: X_i(t) \le \Omega_i\} \quad \text{and} \quad F(t)=\{1,\dots,N\} \setminus J(t).
\end{equation}
The species with indices in $J(t)$ are treated as stochastic, while those with indices in $F(t)$ are treated as deterministic. 

The temporal evolution proceeds as follows:
\begin{itemize}
	\item \textbf{Jump:} Discrete stochastic events affect the components $\mathbf{X}_J$ (simulated exactly using algorithms like the Stochastic Simulation Algorithm), and also impact species in $F$ through reactions involving $J$. Propensities are calculated using the full hybrid state $(\mathbf{X}_J, \mathbf{x}_F)$. Importantly, the propensity usually depends continuously on time due to changing populations in $F$. This is addressed using an event `clock' system \citep{2024arXiv240513239G}.
	\item \textbf{Switch:} Whenever any component crosses its threshold $\Omega_i$ in either direction, the partition sets $J(t)$ and $F(t)$ are instantaneously recomputed, and the governing dynamics for that species immediately switch between stochastic and deterministic treatment.
	\item \textbf{Flow:} Between discrete events, the deterministic components $\mathbf{x}_F$ evolve continuously according to the ODE system Equation~\eqref{eq:ode_limit}, restricted to the species with indices in $F(t)$, and reactions involving only those species. Since $\mathbf{X}_J$ remains constant between jumps and is small by definition, its influence on the flow dynamics can often be neglected.
\end{itemize}

This process is a type of Piecewise-Deterministic Markov Process \citep[PDMP][]{Davis1984}. The dynamics between any two switching events (i.e. for a fixed partition $J(t)$ and $F(t)$) are described by the hybrid generator, $\mathcal{G}_t$. This generator generalises the CTMC generator $\mathcal{L}_t$ by combining discrete jumps with continuous flow. Denoting the combined hybrid state by $\mathbf{z} := (\mathbf{X}_J, \mathbf{x}_F)$, for a function $f$ on the hybrid state space:
\begin{equation}\label{eq:jsf_generator}
	(\mathcal{G}_t f)(\mathbf{z})
	= \sum_{r\in\mathcal{R}_J} a_r(\mathbf{z},t)\big[f(\mathbf{z}+\boldsymbol{\nu}_r)-f(\mathbf{z})\big]
	+ \nabla_{\mathbf{x}_F} f \cdot \sum_{r\in\mathcal{R}_F} \boldsymbol{\nu}_r\, a_r(\mathbf{z},t),
\end{equation}
where the first term captures stochastic jumps and the second term captures deterministic flow. The reaction sets $\mathcal{R}_J$ and $\mathcal{R}_F$ partition the reactions based on which components they affect (for a detailed treatment of this partitioning, see \citealt{2024arXiv240513239G}).


\section{Birth-Death Framework for Extinction Analysis}\label{sec:birth-death-framework}
Before deriving error bounds, we establish the mathematical framework. The JSF method aims to preserve key properties of the exact CTMC. Which properties matter depends on the scientific question under investigation: for instance, one might ask \emph{what is the probability of extinction within a given time horizon?} or \emph{what is the expected species occupancy?} The choice of question defines the objective against which the threshold $\Omega$ should be optimised. Henceforth, we consider a common switching threshold across species and write simply $\Omega$ (i.e., $\Omega_i = \Omega$ for all $i$). If $\Omega$ is set sufficiently large, then the JSF method approaches the CTMC, preserving accuracy but sacrificing computational efficiency. Conversely, while we desire a suitably small $\Omega$ to minimise computational expense, setting $\Omega$ too small risks introducing unacceptable error in the property of interest.

In this work, we focus on the first of these questions: the probability of species extinction within a finite time horizon. We use this as our metric to evaluate the JSF hybrid method's accuracy. The choice is both biologically meaningful and mathematically tractable; extinction events are quintessentially discrete phenomena and cannot be described by continuous models. Our analytical strategy exploits the birth-death structure that emerges near extinction boundaries. When populations are small, the complex non-linear dynamics of the full system reduce to simpler birth-death processes. Specifically, we show how a general continuous-time Markov chain (CTMC) model of a reaction network reduces to a time-inhomogeneous linear birth-death process near extinction, where a subset of species is scarce.
This birth-death framework offers three key advantages. First, it provides exact discrete representation of extinction events. Second, extinction probabilities can be calculated analytically via backward Kolmogorov equations or branching process theory. Third, many population models linearise to birth-death dynamics near absorbing sets, even when their bulk behaviour is non-linear, making this approach broadly applicable.
This reduction enables us to derive theoretical error bounds and practical heuristics for hybrid approximations, providing guidance for threshold selection (e.g., choosing $\Omega$ to trade off accuracy and cost). While we focus on birth-death systems, the underlying mathematical framework extends to more general systems beyond those that are inherently birth-death processes.

\subsection{Linearisation and the Time-Inhomogeneous Birth-Death Process}\label{sec:linearisation}

Our analysis begins by partitioning the system's species into two sets:
\begin{itemize}
	\item \textbf{Scarce species ($J$):} Species $i \in J$ with $X_i \le \Omega$, treated stochastically.
	\item \textbf{Abundant species ($F$):} Species $i \in F$ with $x_i > \Omega$, approximated by deterministic dynamics.
\end{itemize}
This partitioning mirrors the logic of the JSF hybrid methods. The key insight is that when scarce species populations are small, their dynamics linearise. Assuming $\Omega$ is sufficiently small, our analysis relies on the following conditions, which are biologically reasonable near an extinction boundary:
\begin{enumerate}
	\item \textbf{Condition 1 -- Linear reactions dominate:} Reactions involving two or more scarce individuals are rare and their propensities are negligible.
	\item \textbf{Condition 2 -- One-way coupling:} The influence of scarce species on abundant ones is negligible. The abundant species thus evolve independently according to $\dot{\mathbf{x}}_F = f_F(\mathbf{x}_F, t)$, creating a time-varying environment for the scarce species.
	\item \textbf{Condition 3 -- No external input:} For simplicity, we assume no external immigration for scarce species. The framework can be straightforwardly extended to include this as needed.
\end{enumerate}

These conditions constrain the \emph{error-bound derivation}, not the JSF method itself; JSF admits fully bidirectional coupling in simulation. Conditions~1 and~2 are also asymptotically self-consistent near extinction. The neglected nonlinear propensities and the feedback from $\mathbf{X}_J$ to $\mathbf{x}_F$ are both $O(\|\mathbf{X}_J\|)$ at the per-capita level, so they vanish in the regime that controls the extinction probability. The bound is most useful when a small subset of species is rare against a much larger, near-deterministic background: rare-mutant invasion, late-stage viral clearance, pathogen extinction in a host at carrying capacity, minority-species extinction in a large ecosystem, the die-out regime of stochastic SIR dynamics. Sustained feedback from scarce to abundant species, or prolonged near-critical phases where $r(t)\approx 0$, lie outside this analysis. Section~\ref{sec:discussion} returns to these.

In general, the per-capita birth and death rates for the scarce species depend on the full hybrid state $\mathbf{z} = (\mathbf{X}_J, \mathbf{x}_F)$. For each scarce species $i\in J$, define
\begin{itemize}
	\item $\mu_i(\mathbf{z})$: per-capita death rate of type $i$;
	\item $\Lambda_{\ell i}(\mathbf{z})$: per-capita rate at which a parent of type $i$ produces an offspring of type $\ell$.
\end{itemize}

\textit{Condition 1} (linear reactions dominate) says reactions involving two or more scarce individuals are rare. Under standard mass-action scaling, their propensities are $O(\|\mathbf{X}_J\|^2)$, so their \emph{per-capita} contributions are $O(\|\mathbf{X}_J\|)$ and vanish as $\Omega \to 0$. 
\begin{equation}
	\mu_i(\mathbf{z}) \;=\; \mu_i^{(0)}(\mathbf{x}_F) \;+\; O\!\left(\|\mathbf{X}_J\|\right),\qquad
	\Lambda_{\ell i}(\mathbf{z}) \;=\; \Lambda_{\ell i}^{(0)}(\mathbf{x}_F) \;+\; O\!\left(\|\mathbf{X}_J\|\right).
	\label{eq:rate-approx}
\end{equation}
\textit{Condition 2} (one-way coupling) fixes $\mathbf{x}_F(t)$ as a deterministic trajectory. Neglecting the $O(\|\mathbf{X}_J\|)$ terms as $\Omega \to 0$ yields time-dependent per-capita rates that depend only on the abundant-species environment:
\begin{equation}
	\mu_i(t)\;\equiv\;\mu_i^{(0)}\!\big(\mathbf{x}_F(t)\big), \qquad
	\Lambda_{\ell i}(t)\;\equiv\;\Lambda_{\ell i}^{(0)}\!\big(\mathbf{x}_F(t)\big).
\end{equation}
With these rates, each scarce individual evolves independently (dies at rate $\mu_i(t)$; produces an $\ell$-type offspring at rate $\Lambda_{\ell i}(t)$), so $\mathbf{X}_J(t)$ is a time-inhomogeneous multi-type branching (birth–death) process \citep{kendall1948generalized, dewitt2024mean}.

\subsection{Calculating Extinction Probabilities}\label{sec:extinction_probs}

The linearised birth-death structure allows us to compute extinction probabilities using branching process theory. We begin with the general multi-type framework and then specialise to the single-species case, which will be the focus of our subsequent analysis.

Let $\mathbf{q}(s) = (q_i(s))_{i \in J}$ be the vector where $q_i(s)$ is the probability that the lineage (all descendants) of a single individual of
species $i \in J$ present at time $s$ is extinct by the final time $T$. This vector satisfies a backward Kolmogorov system (see Appendix \ref{sec:kolmogorov}):
\begin{equation}
	\frac{d}{ds}\,\mathbf{q}(s)
	= \boldsymbol{\mu}(s) \circ \big(\mathbf{q}(s) - \mathbf{1}\big)
	+ \mathbf{q}(s)\circ\!\Big[\Lambda(s)^{\!\top}\big(\mathbf{1}-\mathbf{q}(s)\big)\Big],
	\qquad \mathbf{q}(T)=\mathbf{0},
	\label{eq:bd-multitype-backward}
\end{equation}
where $\boldsymbol{\mu}(s)$ collects the death rates, $\circ$ denotes the Hadamard (element-wise) product, and the $(\ell,i)$ entry of $\Lambda$ is the per-capita rate at which a parent of type $i$ produces an offspring of type $\ell$. 

The total probability of extinction for a population with integer initial counts $\mathbf n_s=(n_{s,i})_{i\in J}$ at time $s$ and horizon $T$ is then
\begin{equation}
	P_{\text{ext}}(\mathbf n_s, s \mid T) = \prod_{i \in J} q_i(s)^{\,n_{s,i}} .
\end{equation}
In particular, for initial counts $\mathbf n_0=(n_{0,i})_{i\in J}$ at time $s=0$,
\begin{equation}
	P_{\text{ext}}(\mathbf n_0, 0 \mid T) = \prod_{i \in J} q_i(0)^{\,n_{0,i}} .
\end{equation}

While this multi-type formulation is general, for many applications the analysis can be simplified by focusing on a single scarce species, $Y$ (so $J = \{Y\}$). In this scenario, the mathematical framework reduces from matrices and vectors to simple scalars:
\begin{itemize}
	\item The birth matrix $\Lambda(t)$ becomes a $1 \times 1$ matrix, represented by the scalar per-capita birth rate $\lambda(t)$.
	\item The death rate vector $\boldsymbol{\mu}(t)$ becomes a single scalar per-capita death rate $\mu(t)$.
\end{itemize}
As before, these rates, $\lambda(t) \equiv \lambda(\mathbf{x}_F(t))$ and $\mu(t) \equiv \mu(\mathbf{x}_F(t))$, depend on the abundant species' trajectory.

For this single-species case, the extinction probability for a single lineage, denoted $q_Y(s)$, satisfies the scalar backward Riccati equation (forward time $s$):
\begin{equation}
	\frac{dq_Y}{ds} = \mu(s)\big(q_Y - 1\big) + \lambda(s)\big(q_Y - q_Y^2\big),
	\qquad q_Y(T) = 0.
	\label{eq:riccati-extinction}
\end{equation}
Given an initial population $Y(0) = n_0 \in \mathbb{N}$, the total probability of extinction by time $T$ is found by compounding these independent lineage probabilities:
\begin{equation}
	P_{\text{ext}}(n_0, 0 \mid T) = \big[q_Y(0)\big]^{n_0}.
	\label{eq:focal-extinction}
\end{equation}
In Section \ref{sec:error_analysis}, we will use this scalar Riccati equation to derive error bounds; the multi-type framework proceeds analogously via Equation~\eqref{eq:bd-multitype-backward}.

\section{Error analysis for hybrid method}\label{sec:error_analysis}

\subsection{Framing the Analysis: The Collapse Scenario}

To analyse the error of the JSF hybrid method with a given $\Omega$ (recalling that we adopt a common threshold $\Omega$ across species, i.e., $\Omega_i = \Omega$ for all $i$), we first consider the initial state of the focal species $Y$. There are two primary biological scenarios:
\begin{itemize}
	\item \textbf{The introduction scenario}, where a species starts at a low count ($Y(0) < \Omega$), such as during the initial phase of an infection or the arrival of an invasive species.
	\item \textbf{The collapse scenario}, where a previously abundant population declines to the threshold, $Y(s_0) = \Omega$. This represents events such as population crashes or disease clearance.
\end{itemize}

Our analysis will focus on the collapse scenario, which captures the essential error-generating mechanism. To see why, consider that in the introduction scenario, the exact stochastic model and the JSF hybrid model are identical as long as the population remains below the threshold, $Y(t) \le \Omega$. No error can accumulate during this period. If the population hits the threshold at some time $\tau_\Omega < T$, the strong Markov property \citep{oksendal2003stochastic} allows us to restart the analysis from that point. The problem then becomes equivalent to a collapse scenario starting at time $\tau_\Omega$, and error bounds derived for the collapse scenario apply directly, with rates evaluated at the crossing time $\tau_\Omega$.

Focusing on the collapse scenario is powerful because it frames the problem in terms of the immediate dynamics at the boundary. With the population starting at $Y(0) = \Omega$, its subsequent evolution is dictated by the net growth rate, $r(t) = \lambda(t) - \mu(t)$. A negative rate ($r(t) < 0$) creates a ``drift toward extinction,'' tending to keep the population in the stochastic regime, where the models agree. Conversely, a positive rate ($r(t) > 0$) creates a ``drift away from extinction'', pushing the population into the deterministic regime, where approximation errors are generated.

This birth-death perspective, introduced in Section~\ref{sec:birth-death-framework} and here manifest in the interplay between birth rate $\lambda(t)$, death rate $\mu(t)$, and the threshold $\Omega$, is crucial for understanding how and when the JSF approximation introduces significant error, as we develop in the following subsections.

\subsection{Error Metric for Hybrid Approximations}

To quantify the accuracy of the JSF approximation for a given threshold $\Omega$, we define the approximation error as the absolute difference between the total extinction probabilities computed by the exact and hybrid models. We denote the exact probability as $P_{\text{ext}}(\mathbf{x}_0, 0 | T)$ and the JSF approximation as $P_{\text{ext}}^{\rm JSF}(\mathbf{x}_0, 0 | T)$.

Starting from an initial condition $\mathbf{x}_0$ at time $s=0$, the error over a time horizon $T$ is:
\begin{equation}
	\Delta_\Omega(\mathbf{x}_0, T) := \big| P_{\text{ext}}(\mathbf{x}_0, 0 | T) - P_{\text{ext}}^{\rm JSF}(\mathbf{x}_0, 0 | T) \big|.
	\label{eq:delta_def}
\end{equation}
This error metric captures how the choice of threshold $\Omega$ affects the accuracy of extinction probability estimates, providing the foundation for principled threshold selection. When the initial condition is clear from context (e.g., $Y(0) = \Omega$ in the collapse scenario), we write simply $\Delta_\Omega(T)$.

\subsection{Critical Time Regimes and Error Decomposition}

The key insight for our error analysis is that the discrepancy between the exact stochastic model and the JSF hybrid method, $\Delta_\Omega(T)$, arises only when the focal species $Y$ up-crosses the threshold $\Omega$. The first up-crossing time, $\tau_\uparrow$, occurs where $\tau_\uparrow := \inf\{t \geq 0 : Y(t) = \Omega + 1\}$. At this moment, the hybrid model switches to a deterministic ODE, while the true process remains stochastic. The consequences of this divergence depend critically on the system's dynamics at the time of the up-crossing.

The behaviour of the birth-death process is governed by the net growth rate $r(t) = \lambda(t) - \mu(t)$. This allows us to define two crucial time points:
\begin{itemize}
    \item \textbf{The critical time, $t_c$}, is the point where the dynamics shift from being death-dominated ($r(t) < 0$) to birth-dominated ($r(t) \ge 0$). It is defined as:
    \begin{equation}
        t_c := \inf\{t \geq 0 : r(t) \geq 0\}.
    \end{equation}
    Excursions above $\Omega$ before $t_c$ tend to be corrected by the negative drift, whereas those occurring after may persist.
    
    \item \textbf{The point of no return, $s_*$}, represents the latest moment an up-crossing can occur and still be expected to return below $\Omega$ deterministically before the drift turns favourable at $t_c$. It is defined by the solution to (see Appendix \ref{ap:error_bound}):
    \begin{equation}
        \int_{s_*}^{t_c} r(u) \, du = \log\left(\frac{\Omega}{\Omega + 1}\right).
        \label{eq:point-of-no-return}
    \end{equation}
\end{itemize}

\begin{figure}[t]
	\centering
	\includegraphics[width=\linewidth]{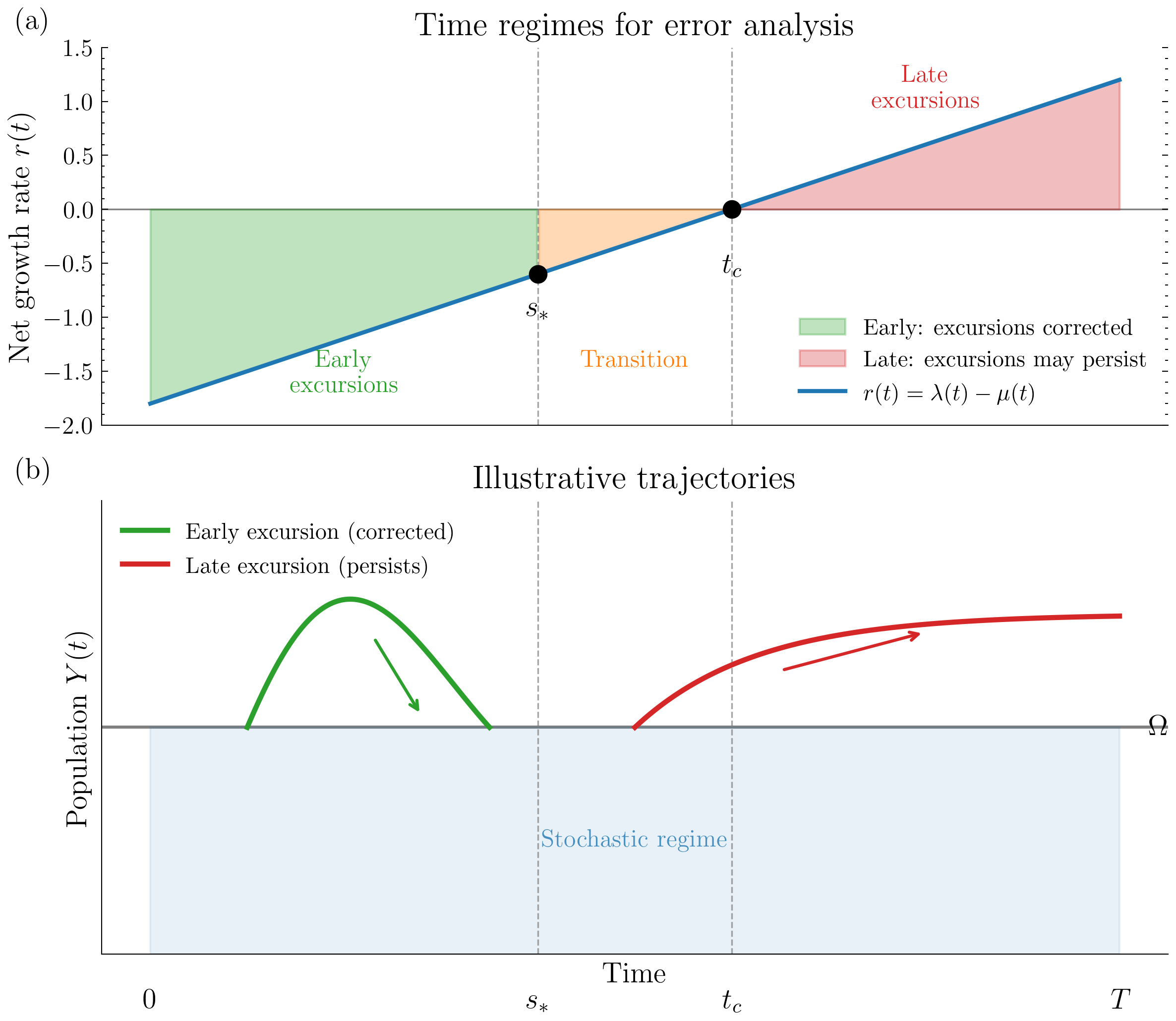}
	\caption{Illustration of the critical time $t_c$ and point of no return $s_*$. (a) The net growth rate $r(t) = \lambda(t) - \mu(t)$ transitions from negative (death-dominated) to positive (birth-dominated) at $t_c$. The point of no return $s_*$ marks the latest time an up-crossing can occur and still be expected to return below $\Omega$ before $t_c$. (b) Illustrative trajectories: early excursions (green) are corrected by negative drift, while late excursions (red) persist due to favourable growth conditions.}
	\label{fig:s_star_diagram}
\end{figure}

These time points, illustrated in Figure~\ref{fig:s_star_diagram}, allow us to partition the total error based on $\tau_\uparrow$. The formal steps for this decomposition are provided in Appendix~\ref{ap:error_bound}. This end result separates the error into two conceptually distinct early/late components:
\begin{equation}
	\Delta_\Omega(T) \le \underbrace{\mathbb{E}\left[ [q_Y(\tau_\uparrow)]^{\Omega+1} \mathbf{1}\{\tau_\uparrow < s_{*}\} \right]}_{\text{Early Excursion Error}} + \underbrace{\mathbb{E}\left[ [q_Y(\tau_\uparrow)]^{\Omega+1} \mathbf{1}\{s_{*} \le \tau_\uparrow < T\} \right]}_{\text{Late Excursion Error}}. \label{eq:early_late_split}
\end{equation}
The subsequent analysis will focus on bounding each of these terms.

\subsection{A rigorous but impractical bound}
A full theoretical bound on the error can be derived by combining the contributions from both early and late excursions. The detailed derivation is presented in Appendix~\ref{ap:error_bound}. In brief, the approach decomposes each stochastic excursion into a deterministic drift component and a martingale fluctuation, bounds the duration of excursions using a drift argument, and then applies the Burkholder--Davis--Gundy (BDG) inequality to control the martingale deviations via their predictable quadratic variation. This yields the following result, cf. Equation~\eqref{eq:early_late_split}:
\begin{equation}
	\Delta_\Omega(T) \le \underbrace{\left(\int_0^{s_*}\Omega\lambda(t)\,dt\right) \frac{L_{\text{eff}}}{\rho} \frac{\Omega+1}{\Omega} C_{\text{BDG}} \sqrt{V_\star}}_{\text{Early Excursion Error}} + \underbrace{\int_{s_*}^T w(s) [q_Y(s)]^{\Omega+1}\,ds}_{\text{Late Excursion Error}}. \label{eq:gen-combined}
\end{equation}
where: $\rho := \inf_{t\in[0,s_*]} \{-r(t)\}$ is the uniform lower bound on the negative drift; $L_{\mathrm{eff}} := \sup_{s\in[0,s_*]} |p'(s)|$ is the Lipschitz constant of the extinction probability $p(t) := q_Y(t)^{\Omega}$; $C_{\mathrm{BDG}}$ is the Burkholder--Davis--Gundy constant; $V_\star := \sup_{s\in[0,s_*]}\int_s^{\sigma(s)} [\lambda(u)+\mu(u)]\,y^{(s)}(u)\,du$ bounds the predictable quadratic variation over early excursions, with $y^{(s)}(u)$ the deterministic trajectory and $\sigma(s)$ its return time to $\Omega$; and $w(s) := \Omega \lambda(s) \exp\bigl(-\Omega \int_0^s \lambda(u) \, du\bigr)$ is the up-crossing time density. See Appendix~\ref{ap:error_bound} for full derivations.

While mathematically rigorous, this bound has significant practical limitations that stem from the early excursion term. First, its implementation is computationally demanding, requiring careful numerical solutions, the evaluation of suprema and infima over time intervals ($L_{\text{eff}}, \rho$), and the estimation of martingale quadratic variations ($V_\star$). Second, and more critically, the bound is often overly conservative, yielding impractically large estimates ($\gg 1$). This is a direct result of compounding several worst-case estimates: the maximum rate of change of the extinction probability ($L_{\text{eff}}$), the weakest possible negative drift ($\rho$), the largest possible stochastic fluctuation ($V_\star$), and the use of a universal $C_{\mathrm{BDG}}$ which controls the worst-case supremum of a martingale from its quadratic variation rather than the typical behaviour. To take a specific example, computing $L_{\mathrm{eff}}$ using a global supremum over the entire early window $[0, s_*]$ overweights periods when the process is far from $\Omega$, even though sensitivity is primarily needed during the rare moments when trajectories remain near the threshold. The use of global bounds on quantities that are only relevant during brief threshold-crossing events leads to a significant overestimation of the error.

Given these limitations, the primary value of the full bound is theoretical. For practical applications, a more direct and computationally tractable approach is needed. This motivates the development of a heuristic that isolates the dominant source of error, which we present in the next section.

\subsection{A Practical Heuristic for Error Estimation}

The rigorous bound derived in the previous section, while theoretically complete, is often computationally intensive and overly conservative for practical use. This motivates a simpler approach that isolates the dominant source of error.

The key insight is that in many systems, the error budget is dominated by the late excursion term. When the death rate significantly exceeds the birth rate in the early phase ($r(t) \ll 0$ for $t < s_*$), any stochastic excursions above the threshold $\Omega$ are corrected rapidly by the strong negative drift. Consequently, the contribution from the early excursion error is often negligible.

This allows us to approximate the total error by retaining only the late term. Bounding the integral for the late excursion probability leads to a simple and effective heuristic (see Appendix~\ref{appendix:heuristic_term} for derivation):
\begin{equation}
	\Delta_\Omega(T) \lesssim [q_Y(s_*)]^{\Omega+1}.
	\label{eq:practical-heuristic}
\end{equation}
We can interpret this formula in a straightforward manner: the dominant error is controlled by the probability of extinction as seen from the ``point of no return'', $s_*$ (see Figure~\ref{fig:s_star_diagram}). This is the latest moment at which a trajectory that has crossed into the deterministic regime can still be expected to return below $\Omega$ before the system's dynamics become favourable for persistent growth.

Algorithm~\ref{alg:heuristic-bound} outlines the straightforward procedure for calculating this heuristic bound.

\begin{algorithm}[H]
	\caption{Calculating the Heuristic Error Bound}
	\label{alg:heuristic-bound}
	\begin{algorithmic}[1]
		\State \textbf{Input:} Threshold $\Omega$, time horizon $T$, birth rate $\lambda(t)$, death rate $\mu(t)$
		\State Compute net growth rate: $r(t) = \lambda(t) - \mu(t)$.
		\State Find the critical time $t_c = \inf\{t \geq 0 : r(t) \geq 0\}$.
		\State Solve for the point of no return $s_*$ using $\int_{s_*}^{t_c} r(u) \, du = \log(\Omega/(\Omega+1))$, cf. Equation~\eqref{eq:point-of-no-return}.
		\State Solve the Riccati equation backward for $q_Y(s)$ on $[s_*, T]$, cf. Equation~\eqref{eq:riccati-extinction}.
		\State \textbf{Evaluate the bound:} $\Delta_\Omega(T) \approx [q_Y(s_*)]^{\Omega+1}$.
		\State \textbf{Output:} Error bound estimate.
	\end{algorithmic}
\end{algorithm}

\section{Case study: The Stochastic Lotka–Volterra Model}

We now demonstrate the practical application of our birth-death framework by analysing a classic predator-prey system with stochastic Lotka-Volterra (LV) dynamics. This system is a suitable case study because it is simple enough to admit explicit analytical solutions while being rich enough to exhibit all the key phenomena of our general theory: time-varying birth-death rates, critical time transitions, and the interplay between stochastic and deterministic dynamics. Moreover, the LV model is sufficiently well-known that readers can focus on the methodology rather than the biological details.

\subsection{The Stochastic Lotka–Volterra Model, scarce prey}\label{sec:lv-case-study}

\subsubsection{Model Definition}

Consider a two-species system with prey $X_1$ (scarce, $J = \{1\}$) and predators $X_2$ (abundant, $F = \{2\}$). The reaction network with rates $\alpha, \beta, \gamma > 0$ is:
\begin{align}
	\text{(Prey birth)} && X_1 &\xrightarrow{\alpha} 2X_1, \\
	\text{(Predation)} && X_1 + X_2 &\xrightarrow{\beta} 2X_2, \\
	\text{(Predator death)} && X_2 &\xrightarrow{\gamma} \emptyset.
\end{align}
When prey is scarce, predation events contribute little to the predator population compared to predator death, which dominates predator dynamics. Under our linearisation assumption (Section~\ref{sec:birth-death-framework}), predator dynamics decouple from prey when prey counts are small, yielding the independent evolution:
\begin{equation}
	\dot{x}_2(t) = -\gamma x_2(t), \qquad \Rightarrow x_2(t) = x_{2,0} e^{-\gamma t}.
	\label{eq:predator-ode}
\end{equation}

This exponential decay of predators directly determines the prey birth-death rates. The prey birth reaction contributes constant rate $\alpha$ to $\lambda(t)$, while predation contributes time-varying rate $\beta x_2(t)$ to $\mu(t)$. Thus (cf. Section \ref{sec:extinction_probs}):
\begin{equation}
	\lambda(t) = \alpha, \qquad \mu(t) = \beta x_{2,0} e^{-\gamma t}, \qquad r(t) = \alpha - \beta x_{2,0} e^{-\gamma t}.
	\label{eq:lv-rates}
\end{equation}

\begin{figure}
	\centering
	\includegraphics[width=\linewidth] {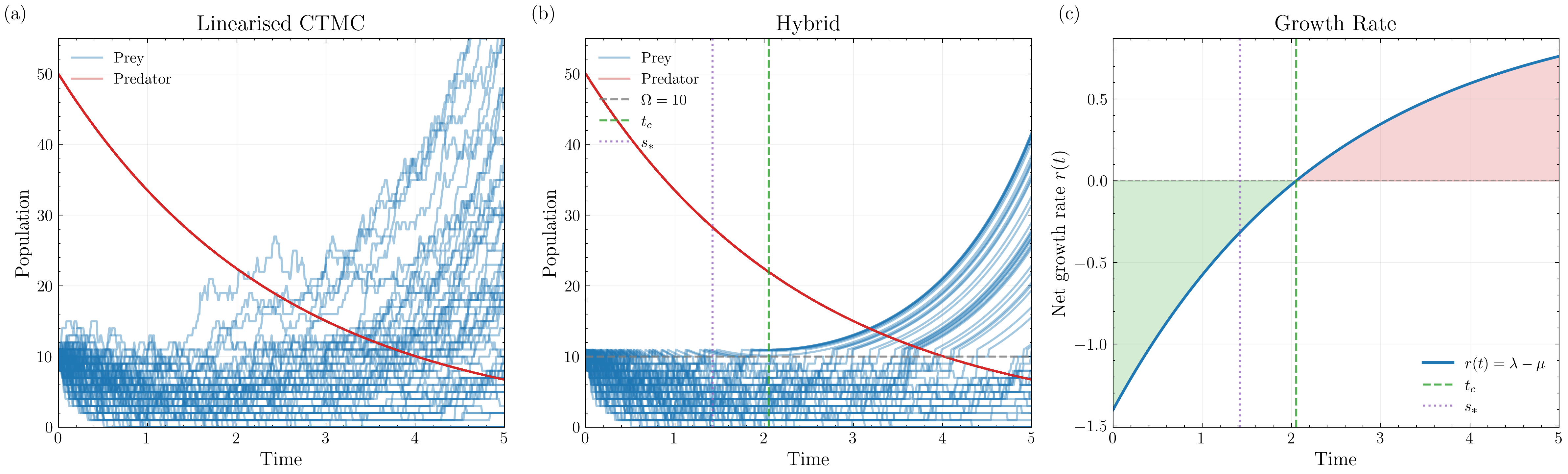}
	\caption{Comparison of the linearised CTMC reference and the JSF hybrid for the Lotka--Volterra predator--prey model. Parameters: $\alpha = 1.10$, $\beta = 0.05$, $\gamma = 0.4$, initial conditions $(x_{1,0}, x_{2,0}) = (10, 50)$. Panel (a) shows $1{,}000$ trajectories of the linearised CTMC reference: prey (blue) evolves stochastically via the SSA and predator (red) follows the deterministic ODE $\dot x_2 = -\gamma x_2$, consistent with the linearisation of Section~\ref{sec:birth-death-framework}. Panel (b) shows $1{,}000$ trajectories of the full JSF hybrid with switching threshold $\Omega=10$: prey is simulated stochastically while $X_1\le\Omega$ and deterministically otherwise, predator as in panel (a). Panel (c) shows the net growth rate $r(t)=\lambda(t)-\mu(t)$, with the critical time $t_c$ (dashed, $r(t_c)=0$) and point of no return $s_*$ (dotted) annotated; the green region is death-dominated ($r<0$) and the red region is birth-dominated ($r>0$).}
	\label{fig:example_system_1}
\end{figure}

\begin{figure}
	\centering
	\includegraphics[width=\linewidth]{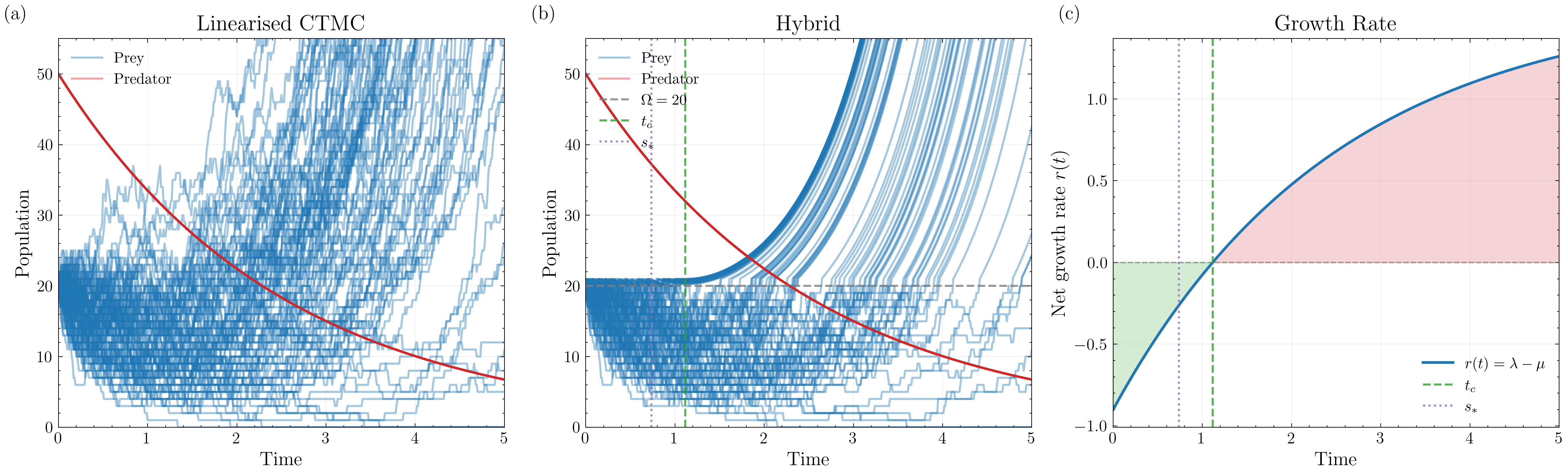}
	\caption{As Fig.~\ref{fig:example_system_1}, with increased prey birth rate $\alpha=1.60$. The higher birth rate lowers extinction probability and shifts both $t_c$ and $s_*$ earlier.}
	\label{fig:example_system_2}
\end{figure}

Figures~\ref{fig:example_system_1}--\ref{fig:example_system_2} show two example simulations of the LV model. We simulate the system under two configurations of the JSF method: a \emph{linearised CTMC reference}, in which the scarce prey is evolved stochastically while the abundant predator follows its deterministic ODE (Equation~\eqref{eq:predator-ode})---consistent with the linearisation assumptions of Section~\ref{sec:birth-death-framework} against which our error bound is derived---and the full \emph{JSF hybrid} with switching threshold $\Omega=10$, in which the prey switches between stochastic and deterministic regimes at $\Omega$. Each configuration aggregates $1{,}000$ realisations of the stochastic process.

\subsubsection{Numerical Validation}\label{sec:results}
All quantities from our general framework — extinction probabilities $q_Y(\cdot)$, critical times $t_c$ and $s_*$, and error bounds — follow by direct substitution of the rates Equation~\eqref{eq:lv-rates} and applying Algorithm \ref{alg:heuristic-bound}. Explicit expressions for the relevant quantities are given in Appendix \ref{ap:lotka_volt}.

To validate our theoretical framework, we conduct Monte Carlo simulations comparing exact stochastic trajectories with JSF hybrid approximations across a range of threshold parameters. This computational experiment serves two purposes: it demonstrates the practical utility of our error bounds and provides empirical validation of the birth-death theoretical predictions.

We consider the Lotka-Volterra system with parameters $\alpha = 1.10$, $\beta = 0.04$, $\gamma = 0.1$, initial conditions $(x_{1,0}, x_{2,0}) = (10, 50)$, and time horizon $T = 7$. For each threshold value $\Omega \in \{5, 10, 15, 20, 25, 30, 35\}$, we perform $32{,}000$ independent simulations using both the linearised CTMC reference (scarce prey simulated by the SSA with the abundant predator evolving deterministically, cf.\ Section~\ref{sec:birth-death-framework}) and the JSF hybrid method. From these simulations, we compute:
\begin{itemize}
	\item \textbf{Empirical error:} $\hat{\Delta}_\Omega(T) = |\hat{P}_{\text{ref}} - \hat{P}_{\text{JSF}}|$, where $\hat{P}$ denotes the empirical extinction probability as determined by Monte Carlo simulation (subscript ``ref'' denotes the linearised CTMC reference)
	\item \textbf{Confidence intervals:} Wilson score intervals accounting for binomial sampling variability
	\item \textbf{Practical heuristic bound:} $[q_Y(s_*)]^{\Omega+1}$ from our late-only approximation (Equation~\ref{eq:practical-heuristic}), computed by solving the Riccati equation and finding the critical time $s_*$ (see Appendix~\ref{ap:lotka_volt})
\end{itemize}

Figure~\ref{fig:error_vs_omega} presents the key results. The empirical error (blue circles with confidence intervals) exhibits the expected monotonic decrease with increasing $\Omega$. As discussed in our theoretical development, when $\Omega$ is small, the hybrid model operates primarily in the deterministic regime, leading to larger discrepancies with the exact stochastic behaviour. As $\Omega$ increases, the models converge since both treat the focal species stochastically throughout most trajectories.

Critically, our practical heuristic bound (red dashed line) provides a reliable upper envelope for the empirical error across all threshold values. This simplified bound, which neglects the computationally intensive early-term contributions and focuses on the dominant late-excursion errors, successfully captures both the magnitude and the functional dependence on $\Omega$. The validation demonstrates that the late-only approximation is indeed sufficient for practical threshold selection in this birth-death system.

\begin{figure}
	\centering
		\includegraphics[width=\linewidth]{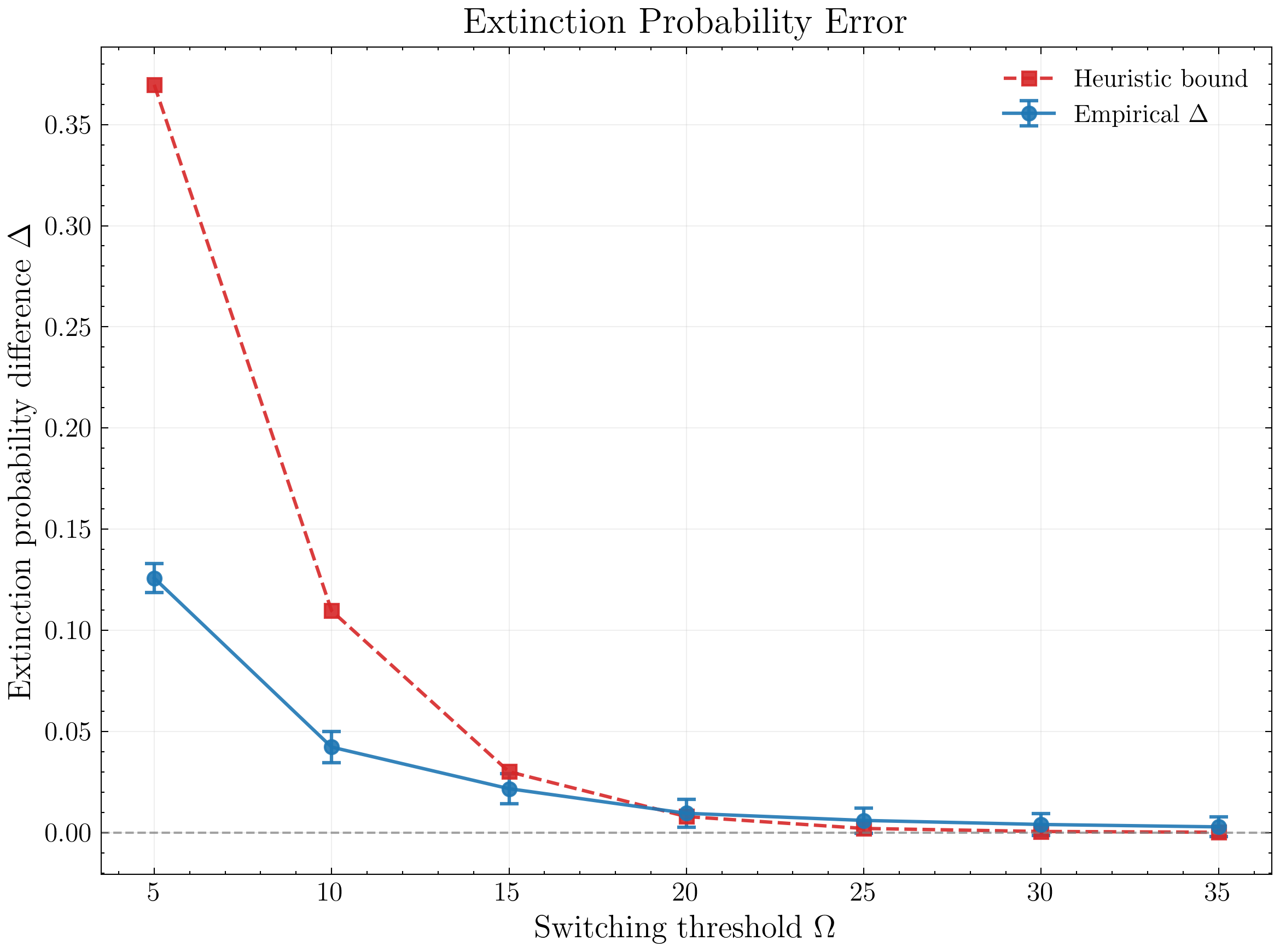}
	\caption{Validation of the practical heuristic bound for the Lotka-Volterra system. Blue circles show empirical error $\Delta_\Omega(T)$ between the linearised CTMC reference (scarce prey via SSA, abundant predator via ODE, cf.\ Section~\ref{sec:birth-death-framework}) and the JSF hybrid simulations ($32{,}000$ replications each), with Wilson score confidence intervals. Red dashed line shows the late-only heuristic bound $[q_Y(s_*)]^{\Omega+1}$. The simplified bound provides a reliable upper envelope, demonstrating the effectiveness of the practical approximation. Parameters: $\alpha = 1.10$, $\beta = 0.04$, $\gamma = 0.1$, $(x_{1,0}, x_{2,0}) = (\Omega, 40)$, $T = 7$.}
	\label{fig:error_vs_omega}
\end{figure}

\subsubsection{Parameter Space Exploration}\label{sec:parameter-space}

  The previous analysis focused on a particular instance of the Lotka-Volterra system with fixed parameters. To assess the robustness of our theoretical framework and
the practical heuristic bound across different dynamical regimes, we now systematically explore the parameter space.

We vary the prey birth rate $\alpha \in [1.0, 1.2]$ and predation rate $\beta \in [0.04, 0.06]$ while holding the predator death rate fixed at $\gamma = 0.4$, with
initial conditions $(x_{1,0}, x_{2,0}) = (\Omega, 40)$ and time horizon $T = 7$. This parameter range spans systems from near-certain extinction (low $\alpha$, high
$\beta$) to high survival probability (high $\alpha$, low $\beta$). For each of the 36 parameter combinations on a $6 \times 6$ grid, we compute the heuristic
monotonicity bound for $\Omega = 10$ and perform $32{,}000$ independent simulations each of the linearised CTMC reference and JSF hybrid methods. We calculate the empirical extinction
probability gap $\Delta_{\text{emp}} = p_{\text{stochastic}} - p_{\text{hybrid}}$ and compare it to the deterministic theoretical bound, computing the margin (bound
$-$ empirical) and its statistical significance via z-score (margin / standard error).

\begin{figure}
	\centering
	\includegraphics[width=0.8\linewidth]{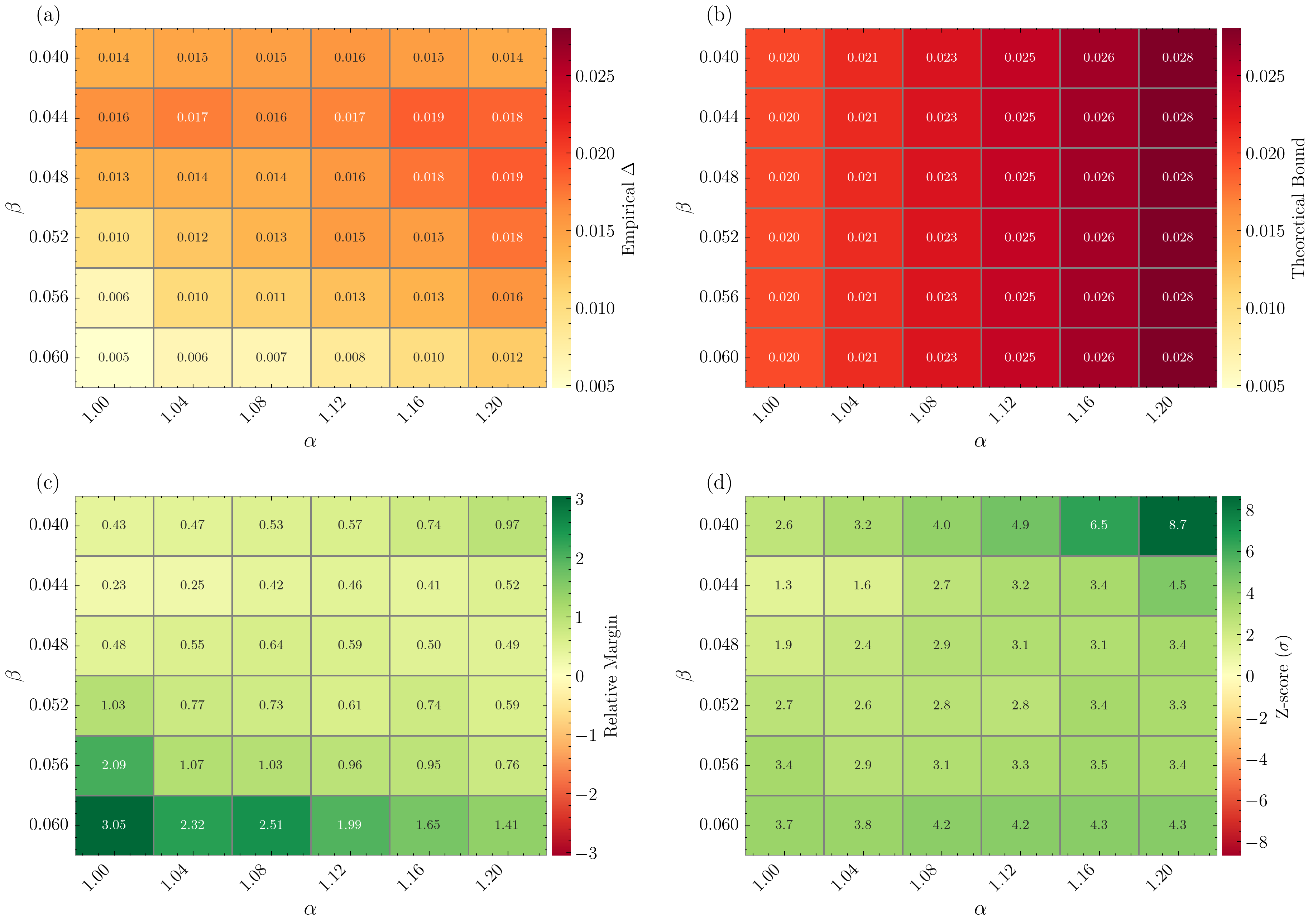}
	\caption{Parameter space exploration for $\Omega=10$ across prey birth rate $\alpha$ and predation rate $\beta$. (a) Empirical extinction probability gap
	$\Delta_{\text{emp}}$ from $32{,}000$ simulations per grid point. (b) Deterministic theoretical bound values showing monotonic dependence on $\alpha$ but near-invariance
	to $\beta$. (c) Margin (bound slack) with green indicating positive margins; the black dashed contour at margin=0 indicates the bound-violation boundary.
	(d) Statistical significance (z-score = margin/error) with green indicating the bound holds and red indicating violations. All 36 parameter combinations show positive margins, with z-scores ranging from $1.3\sigma$ to $8.7\sigma$, validating the theoretical framework across diverse dynamical regimes
	from high-extinction to high-survival scenarios.}
	\label{fig:parameter_heatmap}
\end{figure}

Figure~\ref{fig:parameter_heatmap} presents the results across four complementary views of the parameter space. The practical heuristic bound holds reliably across
the parameter range tested, with margins ranging from $0.004$ to $0.016$ and z-scores from $1.3\sigma$ to $8.7\sigma$. All 36 grid points exhibit strictly positive margins, and every z-score is at least $1.3\sigma$ above zero, providing strong empirical support for the bound across the tested parameter range. While the theoretical bound is deterministic, the empirical gap has sampling
uncertainty from finite simulations (Bernoulli trials), making the z-score essential for assessing whether observed margins reflect true bound slack versus
measurement noise.

Notably, the theoretical bound exhibits near-perfect linear growth with increasing prey birth rate $\alpha$ (correlation of $0.9998$, from $\approx 0.020$ at $\alpha=1.0$ to $\approx 0.028$ at
$\alpha=1.2$), reflecting increased coupling strength between the exact and hybrid dynamics in more favourable demographic conditions. This dependence on $\alpha$ can be understood practically: larger prey birth rates increase the net growth rate $r = \alpha - \beta x_2$ of the prey population, amplifying the rate at which stochastic and deterministic trajectories can diverge before reaching the extinction boundary. In contrast, the bound remains
nearly invariant to changes in predation rate $\beta$, varying by less than $0.001$ within each $\alpha$ slice. The empirical gap shows similar trends but with
greater sensitivity to $\beta$, resulting in margin variation across both dimensions. The tightest margins occur at low $\alpha=1.0$ where the bound approaches the empirical gap most closely, while the loosest margins ($\approx 0.016$) appear at intermediate parameter values where the empirical gap is smallest relative to the bound.

%
%
%
%

\section{Discussion and Conclusions}\label{sec:discussion}

We have demonstrated that principled threshold selection for JSF-type hybrid methods can be achieved through birth-death analysis near extinction boundaries. Hybrid multiscale schemes, like JSF, promise CTMC fidelity where discreteness matters and ODE speed where it does not, but the practical difficulty in deployment (i.e. choosing the switching threshold(s) $\Omega$) has remained largely arbitrary. Our framework reframes this challenge as an extinction-accuracy budgeting problem with explicit, computable bounds.

By linearizing near absorbing sets and exploiting birth-death structure, we separated the approximation error into two components: (i) \emph{early excursions} under negative drift ($r(t)<0$), which are typically self-correcting and contribute minimally to the error budget, and (ii) \emph{late excursions} once the net growth becomes nonnegative ($r(t)\geq 0$), which dominate bias in extinction observables. This decomposition yields both a rigorous theoretical bound and, crucially, a compact heuristic that requires only solving a single Riccati equation rather than thousands of Monte Carlo simulations.

Our validation with the case study involving the Lotka-Volterra system confirms that the late-only estimate, Equation~\eqref{eq:practical-heuristic} reliably captures the main failure mode of JSF for extinction problems. This bound identifies paths that cross into the deterministic regime near or after the critical time $t_c$ when $r(t)$ changes sign, and thus do not return quickly to the stochastic boundary where discreteness matters. When there is  a distinct period of death dominance on $[0,s_*]$ followed by a transition to birth dominance, early excursions are rapidly corrected by negative drift, and the late term controls the bias. In order to choose a value of $\Omega$ a user must fix a time horizon $T$ and an extinction-probability error tolerance $\varepsilon$, and choose
\begin{equation}
\Omega \geq \left\lceil \frac{\log \varepsilon}{\log q_Y(s_*)} \right\rceil - 1,
\end{equation}
	The heuristic approach scales as $O(T)$ for solving the Riccati equation \citep{LaineTomlin2018ParallelRiccati} plus $O(\log 1/\delta)$ for the one-dimensional root finding required to determine $s_\star$, where $\delta$ is the numerical tolerance used in the bisection method \citep{BurdenFairesBurden2016}, see Appendix \ref{ap:getting_s_star}. This contrasts with $O(N \times T \times R)$ for Monte Carlo validation with $N$ replications, $T$ time steps, and $R$ reactions per step.

For multiple scarce species $i \in J$, a union bound over species-specific errors gives
\begin{equation}
\Delta_\Omega(T) \lesssim \sum_{i \in J} [q_i(s_{*,i})]^{\Omega_i+1}.
\end{equation}
This bound holds because the total error cannot exceed the sum of individual species errors. As before, to specify switching thresholds $\Omega_i$, first allocate per-species error budgets $\varepsilon_i$ with $\sum_i \varepsilon_i \leq \varepsilon$ and then choose the thresholds $\Omega_i$ accordingly. In practice, it is recommended to start with equal budgets and then tighten those for species with large $q_i(s_{*,i})$ values.

\subsection{Limitations and scope of applicability}

The heuristic relies on three modelling assumptions that underpin the linearisation (Section~\ref{sec:linearisation}): scarce--scarce interactions are $O(\|\mathbf{X}_J\|^2)$ and so negligible near extinction; the coupling from abundant to scarce species dominates the reverse coupling in the low-count regime; and scarce species receive no external immigration (or immigration is modelled explicitly). These bind the error-bound derivation, not the JSF simulation: JSF is indifferent to the coupling structure, and the assumptions are simply the minimal conditions under which the scarce-species dynamics near extinction reduce to a time-inhomogeneous birth--death process with a tractable Riccati equation. The first and second are asymptotically self-consistent. Both neglected contributions are $O(\|\mathbf{X}_J\|)$ at the per-capita level and vanish in the regime that determines extinction. The framework then applies to systems in which a rare subset lives inside a large, near-deterministic background: rare-mutant invasion, late-stage viral clearance, pathogen extinction in a host at carrying capacity, SIR die-out.

Even within scope, the heuristic's tightness depends on whether the dynamics separate cleanly into a death-dominated early phase and a birth-dominated late phase. The Lotka--Volterra case study (Section~\ref{sec:lv-case-study}) sits in this regime: the one-way approximation enters when we drop the predation contribution to predator dynamics in Equation~\eqref{eq:predator-ode}, leaving predators to decay exponentially. The dropped term is $O(X_1)$, so this is justified when prey are scarce, but it would need to be retained in systems where scarce prey sustain an otherwise weakly-coupled predator population, or more generally where there is strong feedback from $\mathbf{X}_J$ onto $\mathbf{x}_F$.

The heuristic can also under-estimate the error in two further regimes. If the system has a sustained near-critical phase where $r(t)\approx 0$, early excursions are no longer rapidly corrected and the early term in the rigorous bound (Equation~\eqref{eq:gen-combined}) cannot be neglected. If scarce--scarce interactions are strong (e.g., cooperative effects, Allee dynamics), neglecting the $O(\|\mathbf{X}_J\|^2)$ propensities changes the extinction dynamics qualitatively rather than only quantitatively. In both cases, use the full bound in Equation~\eqref{eq:gen-combined} or apply a conservative margin to the heuristic.

\subsection{Broader implications and future directions}

The error-control framework is method-agnostic. The heuristic bound $[q_Y(s_*)]^{\Omega+1}$ depends only on the birth and death rates $\lambda(t),\mu(t)$ of the scarce species near the extinction boundary, the time horizon $T$, and the threshold $\Omega$. It does not depend on the rule by which a hybrid simulator decides when to switch between stochastic and deterministic regimes, nor on how the continuous component is integrated between jumps. The same inequality therefore applies, essentially unchanged, to the multi-level schemes of \citet{WinkelmannSchutte2017,WinkelmannSchutte2020}, the regime-conversion method of \citet{Kynaston2023}, partitioned leaping \citep{Cao2005Partition}, and the PDMP-style hybrids of \citet{Crudu2009Hybrid,Marchetti2016HRSSA,Alfonsi2005Adaptive}, provided the simulator preserves the stochastic birth--death dynamics of the scarce species below the threshold, as all threshold-based hybrids do by construction. JSF is the simulator we use here, but the analysis is not tied to it: the contribution is a recipe for equipping any threshold-based hybrid with an \emph{a priori} error bound on extinction observables.

Beyond the choice of simulator, the birth--death reduction is a template for accuracy control in any rare-event problem where linearisation near a critical point is valid. Three extensions follow naturally. Time- or state-dependent thresholds $\Omega(t)$ that grow as $t \to t_c$ would suppress late-excursion bias. Modified backward equations would carry the analysis to first-passage times, outbreak probabilities, and other terminal functionals. The same construction should also adapt to non-extinction observables such as moments and steady-state distributions.

In practice the model parameters $\boldsymbol\theta$ are often unknown and inferred from data, while our threshold rule uses specific parameter values to compute $s_*(\boldsymbol\theta)$ and $q_Y(s_*;\boldsymbol\theta)$. This is appropriate, since the correct treatment of the system depends on $\boldsymbol\theta$, but it couples inference and hybridisation: the choice of $\Omega$ affects computational cost, and a too-small $\Omega$ can bias the likelihood or summary statistics used for calibration. We leave this for future work.


\section*{Data Availability Statement}
Data sharing is not applicable to this article as no datasets were generated or analysed during the current study.

\bibliographystyle{plainnat}
\bibliography{references}  

@article{dewitt2024mean,
  title={Mean-field interacting multi-type birth--death processes with a view to applications in phylodynamics},
  author={DeWitt, William S and Evans, Steven N and Hiesmayr, Ella and Hummel, Sebastian},
  journal={Theoretical Population Biology},
  volume={159},
  pages={1--12},
  year={2024},
  publisher={Elsevier}
}

@article{kendall1948generalized,
  title={On the generalized" birth-and-death" process},
  author={Kendall, David G},
  journal={The annals of mathematical statistics},
  pages={1--15},
  year={1948},
  publisher={JSTOR}
}

@article{Doran2023,
	title = {Mathematical methods for scaling from within-host to population-scale in infectious disease systems},
	journal = {Epidemics},
	volume = {45},
	pages = {100724},
	year = {2023},
	issn = {1755-4365},
	doi = {https://doi.org/10.1016/j.epidem.2023.100724},
	url = {https://www.sciencedirect.com/science/article/pii/S1755436523000609},
	author = {James W.G. Doran and Robin N. Thompson and Christian A. Yates and Ruth Bowness}
}

@article{Childs2019,
	author       = {Childs, Lauren M. and El Moustaid, Fadoua and Gajewski, Zachary and Kadelka, Sarah and Nikin-Beers, Ryan and Smith, John W. Jr. and Walker, Melody and Johnson, Leah R.},
	title        = {Linked within‐host and between‐host models and data for infectious diseases: a systematic review},
	journal      = {PeerJ},
	year         = {2019},
	month        = {6},
	volume       = {7},
	pages        = {e7057},
	doi          = {10.7717/peerj.7057}
}

@article{Wang2022,
	author = {Wang, Xueying and Wang, Sunpeng and Wang, Jin and Rong, Libin},
	year = {2022},
	month = {09},
	pages = {},
	title = {A Multiscale Model of COVID-19 Dynamics},
	volume = {84},
	journal = {Bulletin of Mathematical Biology},
	doi = {10.1007/s11538-022-01058-8}
}

@article{Ludwig1978,
	ISSN = {00218790, 13652656},
	URL = {http://www.jstor.org/stable/3939},
	author = {Donald. Ludwig and Dixon D. Jones and Crawford S. Holling},
	journal = {Journal of Animal Ecology},
	number = {1},
	pages = {315--332},
	publisher = {[Wiley, British Ecological Society]},
	title = {Qualitative Analysis of Insect Outbreak Systems: The Spruce Budworm and Forest},
	urldate = {2025-09-04},
	volume = {47},
	year = {1978}
}

@article{Levin1992,
	author = {Levin, Simon A.},
	title = {The Problem of Pattern and Scale in Ecology: The Robert H. MacArthur Award Lecture},
	journal = {Ecology},
	volume = {73},
	number = {6},
	pages = {1943-1967},
	doi = {https://doi.org/10.2307/1941447},
	url = {https://esajournals.onlinelibrary.wiley.com/doi/abs/10.2307/1941447},
	eprint = {https://esajournals.onlinelibrary.wiley.com/doi/pdf/10.2307/1941447},
	year = {1992}
}

@article{Nichols2008,
	ISSN = {00218901, 13652664},
	URL = {http://www.jstor.org/stable/20144099},
	author = {James D. Nichols and Larissa L. Bailey and Allan F. O'Connell and Neil W. Talancy and Evan H. Campbell Grant and Andrew T. Gilbert and Elizabeth M. Annand and Thomas P. Husband and James E. Hines},
	journal = {Journal of Applied Ecology},
	number = {5},
	pages = {1321--1329},
	publisher = {[British Ecological Society, Wiley]},
	title = {Multi-Scale Occupancy Estimation and Modelling Using Multiple Detection Methods},
	urldate = {2025-09-04},
	volume = {45},
	year = {2008}
}

@article{McGarigal_2016, title={Multi-scale habitat selection modeling: a review and outlook}, volume={31}, ISSN={1572-9761}, url={http://dx.doi.org/10.1007/s10980-016-0374-x}, DOI={10.1007/s10980-016-0374-x}, number={6}, journal={Landscape Ecology}, publisher={Springer Science and Business Media LLC}, author={McGarigal, Kevin and Wan, Ho Yi and Zeller, Kathy A. and Timm, Brad C. and Cushman, Samuel A.}, year={2016}, month=apr, pages={1161–1175} }

@article{LaineTomlin2018ParallelRiccati,
  title   = {The Parallelization of Riccati Recursion},
  author  = {Laine, Forrest and Tomlin, Claire},
  journal = {arXiv preprint arXiv:1809.06360},
  year    = {2018},
  url     = {https://arxiv.org/abs/1809.06360}
}

@book{andersonkurtz,
author = {Anderson, David F. and Kurtz, Thomas G.},
title = {Stochastic Analysis of Biochemical Systems},
year = {2015},
isbn = {3319168940},
publisher = {Springer Publishing Company, Incorporated}
}

@book{BurdenFairesBurden2016,
  title     = {Numerical Analysis},
  author    = {Burden, Richard L. and Faires, J. Douglas and Burden, Annette M.},
  edition   = {10th},
  year      = {2016},
  publisher = {Cengage},
}

@article{Dada2011,
	author  = {Dada, Joseph O. and Mendes, Pedro},
	title   = {Multi-scale modelling and simulation in systems biology},
	journal = {Integrative Biology},
	year    = {2011},
	volume  = {3},
	number  = {2},
	pages   = {86--96},
	doi     = {10.1039/c0ib00075b}
}

@article{Walpole2013,
	author  = {Walpole, Joseph and Papin, Jason A. and Peirce, Shayn M.},
	title   = {Multiscale Computational Models of Complex Biological Systems},
	journal = {Annual Review of Biomedical Engineering},
	year    = {2013},
	volume  = {15},
	pages   = {137--154},
	doi     = {10.1146/annurev-bioeng-071811-150104}
}

@article{Qu2011,
	author  = {Qu, Zhilin and Garfinkel, Alan and Weiss, James N. and Nivala, Melissa},
	title   = {Multi-scale modeling in biology: How to bridge the gaps between scales?},
	journal = {Progress in Biophysics and Molecular Biology},
	year    = {2011},
	volume  = {107},
	number  = {1},
	pages   = {21--31},
	doi     = {10.1016/j.pbiomolbio.2011.06.004}
}

@article{kurtz1970solutions,
	title        = {Solutions of ordinary differential equations as limits of pure jump Markov processes},
	author       = {Kurtz, Thomas G.},
	journal      = {Journal of Applied Probability},
	volume       = {7},
	number       = {1},
	pages        = {49--58},
	year         = {1970},
	publisher    = {Applied Probability Trust}
}

@article{kurtz1971limit,
	title        = {Limit theorems for sequences of jump Markov processes approximating ordinary differential processes},
	author       = {Kurtz, Thomas G.},
	journal      = {Journal of Applied Probability},
	volume       = {8},
	number       = {2},
	pages        = {344--356},
	year         = {1971},
	publisher    = {Applied Probability Trust}
}

@article{kurtz1972relationship,
	title        = {The relationship between stochastic and deterministic models for chemical reactions},
	author       = {Kurtz, Thomas G.},
	journal      = {Journal of Chemical Physics},
	volume       = {57},
	number       = {7},
	pages        = {2976--2978},
	year         = {1972},
	publisher    = {American Institute of Physics}
}

@article{Fowler2021AttoFoxes,
	author       = {Fowler, Andrew C.},
	title        = {Atto-Foxes and Other Minutiae},
	journal      = {Bulletin of Mathematical Biology},
	year         = {2021},
	volume       = {83},
	number       = {10},
	pages        = {Article 104},
	doi          = {10.1007/s11538-021-00936-x},
}

@article{LobrySari2015,
	author  = {Lobry, Claude and Sari, Tewfik},
	title   = {Migrations in the Rosenzweig--MacArthur model and the "atto-fox" problem},
	journal = {ARIMA Journal},
	year    = {2015},
	volume  = {20},
	pages   = {95--125},
	doi     = {10.46298/arima.1990},
}

@article{Doob1942TopicsMarkoffChains,
	author  = {Doob, Joseph L.},
	title   = {Topics in the Theory of Markoff Chains},
	journal = {Transactions of the American Mathematical Society},
	year    = {1942},
	volume  = {52},
	number  = {1},
	pages   = {37--64},
	doi     = {10.1090/S0002-9947-1942-0006633-7},
}

@article{Doob1945MarkoffChainsDenumerable,
	author  = {Doob, Joseph L.},
	title   = {Markoff Chains---Denumerable Case},
	journal = {Transactions of the American Mathematical Society},
	year    = {1945},
	volume  = {58},
	number  = {3},
	pages   = {455--473},
	doi     = {10.1090/S0002-9947-1945-0013857-4},
}

@article{Gillespie1976GeneralMethodSSA,
	author  = {Gillespie, Daniel T.},
	title   = {A General Method for Numerically Simulating the Stochastic Time Evolution of Coupled Chemical Reactions},
	journal = {Journal of Computational Physics},
	year    = {1976},
	volume  = {22},
	number  = {4},
	pages   = {403--434},
	doi     = {10.1016/0021-9991(76)90041-3},
}

@article{Gillespie1977ExactSSA,
	author  = {Gillespie, Daniel T.},
	title   = {Exact Stochastic Simulation of Coupled Chemical Reactions},
	journal = {The Journal of Physical Chemistry},
	year    = {1977},
	volume  = {81},
	number  = {25},
	pages   = {2340--2361},
	doi     = {10.1021/j100540a008},
}

@article{Gillespie2001TauLeaping,
	author = {Gillespie, Daniel T.},
	title = {Approximate accelerated stochastic simulation of chemically reacting systems},
	journal = {The Journal of Chemical Physics},
	year = {2001},
	volume = {115},
	number = {4},
	pages = {1716--1733},
	doi = {10.1063/1.1378322}
}

@article{Gillespie2000CLE,
	author = {Gillespie, Daniel T.},
	title = {The chemical Langevin equation},
	journal = {The Journal of Chemical Physics},
	year = {2000},
	volume = {113},
	number = {1},
	pages = {297--306},
	doi = {10.1063/1.481811}
}

@article{Rao2003CLE,
	author = {Rao, Christopher V. and Arkin, Adam P.},
	title = {Stochastic chemical kinetics and the quasi-steady-state assumption: Application to the Gillespie algorithm},
	journal = {The Journal of Chemical Physics},
	year = {2003},
	volume = {118},
	number = {11},
	pages = {4999--5010},
	doi = {10.1063/1.1545446}
}

@article{Cao2006CLE,
	author = {Cao, Yang and Gillespie, Daniel T. and Petzold, Linda R.},
	title = {Efficient step size selection for the tau-leaping simulation method},
	journal = {The Journal of Chemical Physics},
	year = {2006},
	volume = {124},
	number = {4},
	pages = {044109},
	doi = {10.1063/1.2159468}
}

@article{Gibson2000CLE,
	author = {Gibson, Michael A. and Bruck, Jehoshua},
	title = {Efficient exact stochastic simulation of chemical systems with many species and many channels},
	journal = {The Journal of Physical Chemistry A},
	year = {2000},
	volume = {104},
	number = {9},
	pages = {1876--1889},
	doi = {10.1021/jp993732q}
}

@article{Simoni2019SSAReview,
	author       = {Simoni, Giulia and Reali, Federico and Priami, Corrado and Marchetti, Luca},
	title        = {Stochastic simulation algorithms for computational systems biology: Exact, approximate, and hybrid methods},
	journal      = {Wiley Interdisciplinary Reviews: Systems Biology and Medicine},
	year         = {2019},
	volume       = {11},
	number       = {6},
	pages        = {e1459},
	doi          = {10.1002/wsbm.1459}
}

@article{cotter2016error,
	author       = {Cotter, Simon L. and Erban, Radek},
	title        = {Error Analysis of Diffusion Approximation Methods for Multiscale Systems in Reaction Kinetics},
	journal      = {SIAM Journal on Scientific Computing},
	year         = {2016},
	volume       = {38},
	number       = {1},
	pages        = {B144--B163},
	doi          = {10.1137/14100052X}
}

@article{flegg2014analysis,
	author       = {Flegg, Mark B. and Chapman, S. Jonathan and Zheng, Lei and Erban, Radek},
	title        = {Analysis of the Two-Regime Method on Square Meshes},
	journal      = {SIAM Journal on Scientific Computing},
	year         = {2014},
	volume       = {36},
	number       = {3},
	pages        = {B561--B588},
	doi          = {10.1137/130915844}
}

@article{isaacson2013convergent,
	author       = {Isaacson, Samuel A.},
	title        = {A convergent reaction-diffusion master equation},
	journal      = {Journal of Chemical Physics},
	year         = {2013},
	volume       = {139},
	number       = {5},
	pages        = {054101},
	doi          = {10.1063/1.4816377}
}

@article{Haseltine2002,
	author  = {Haseltine, Eric L. and Rawlings, James B.},
	title   = {Approximate Simulation of Coupled Fast and Slow Reactions for Stochastic Chemical Kinetics},
	journal = {Journal of Chemical Physics},
	year    = {2002},
	volume  = {117},
	number  = {15},
	pages   = {6959--6969},
	doi     = {10.1063/1.1505860},
}

@article{Cao2005Partition,
	author  = {Cao, Yang and Gillespie, Daniel T. and Petzold, Linda R.},
	title   = {Multiscale Stochastic Simulation Algorithm with Stochastic Partial Equilibrium Assumptions for Chemically Reacting Systems},
	journal = {Journal of Computational Physics},
	year    = {2005},
	volume  = {206},
	number  = {2},
	pages   = {395--411},
	doi     = {10.1016/j.jcp.2004.12.014},
}

@article{Buckwar2011,
  author  = {Buckwar, Evelyn and Riedler, Martin G.},
  title   = {Runge--Kutta methods for jump--diffusion differential equations},
  journal = {Journal of Computational and Applied Mathematics},
  year    = {2011},
  volume  = {236},
  number  = {6},
  pages   = {1155--1182},
  doi     = {10.1016/j.cam.2011.08.001}
}

@book{Walter1998,
	author    = {Walter, Wolfgang},
	title     = {Ordinary Differential Equations},
	publisher = {Springer},
	year      = {1998},
	series    = {Graduate Texts in Mathematics},
	doi       = {10.1007/978-1-4612-0601-9}
}

@article{Gronwall1919,
  author       = {Gr{\"o}nwall, Thomas Hakon},
  title        = {Note on the derivatives with respect to a parameter of the solutions of a system of differential equations},
  journal      = {Annals of Mathematics},
  volume       = {20},
  number       = {4},
  pages        = {292--296},
  year         = {1919},
  publisher    = {JSTOR}
}

@article{Burkholder1966,
  author  = {Burkholder, D. L.},
  title   = {Martingale transforms},
  journal = {Annals of Mathematical Statistics},
  volume  = {37},
  number  = {6},
  pages   = {1494--1504},
  year    = {1966}
}

@book{oksendal2003stochastic,
	title={Stochastic Differential Equations: An Introduction with Applications},
	author={{\O}ksendal, Bernt},
	edition={6th},
	year={2003},
	publisher={Springer},
	address={Berlin},
	doi={10.1007/978-3-642-14394-6}
}

@article{Davis1970,
  author  = {Davis, Burgess},
  title   = {On the integrability of the martingale square function},
  journal = {Israel Journal of Mathematics},
  volume  = {8},
  number  = {2},
  pages   = {187--190},
  year    = {1970}
}

@article{BurkholderGundy1970,
  author  = {Burkholder, D. L. and Gundy, R. F.},
  title   = {Extrapolation and Interpolation of Quasi-Linear Operators on Martingales},
  journal = {Acta Mathematica},
  volume  = {124},
  pages   = {249--304},
  year    = {1970}
}

@article{ANGIUS201549,
	title = {Approximate analysis of biological systems by hybrid switching jump diffusion},
	journal = {Theoretical Computer Science},
	volume = {587},
	pages = {49-72},
	year = {2015},
	note = {Interactions between Computer Science and Biology},
	issn = {0304-3975},
	doi = {https://doi.org/10.1016/j.tcs.2015.03.015},
	url = {https://www.sciencedirect.com/science/article/pii/S0304397515002194},
	author = {Alessio Angius and Gianfranco Balbo and Marco Beccuti and Enrico Bibbona and Andras Horvath and Roberta Sirovich}
}

@inproceedings{Alfonsi2005Adaptive,
	author    = {Alfonsi, Aur\'elien and Canc\`es, Eric and Turinici, Gabriel and
	Di Ventura, Barbara and Huisinga, Wilhelm},
	title     = {Adaptive Simulation of Hybrid Stochastic and Deterministic
	Models for Biochemical Systems},
	booktitle = {ESAIM: Proceedings (CEMRACS 2004 -- Mathematics and Applications
	to Biology and Medicine)},
	year      = {2005},
	volume    = {14},
	pages     = {1--13},
	note      = {Presented at CEMRACS 2004},
	doi       = {10.1051/proc:2005001},
}

@article{Marchetti2016HRSSA,
	author  = {Marchetti, Luca and Priami, Corrado and Thanh, Vo Hong},
	title   = {HRSSA – Efficient hybrid stochastic simulation for spatially homogeneous biochemical reaction networks},
	journal = {Journal of Computational Physics},
	year    = {2016},
	volume  = {317},
	pages   = {301--317},
	doi     = {10.1016/j.jcp.2016.04.056},
}

@article{Crudu2009Hybrid,
	author  = {Crudu, Alina and Debussche, Arnaud and Radulescu, Ovidiu},
	title   = {Hybrid stochastic simplifications for multiscale gene networks},
	journal = {BMC Systems Biology},
	year    = {2009},
	volume  = {3},
	number  = {89},
	doi     = {10.1186/1752-0509-3-89},
}

@ARTICLE{2024arXiv240513239G,
	author = {{Germano}, Domenic P.~J. and {Zarebski}, Alexander E. and {Hautphenne}, Sophie and {Moss}, Robert and {Flegg}, Jennifer A. and {Flegg}, Mark B.},
	title = "{A hybrid framework for compartmental models enabling simulation-based inference}",
	journal = {arXiv e-prints},
	year = 2024,
	month = may,
	eid = {arXiv:2405.13239},
	pages = {arXiv:2405.13239},
	doi = {10.48550/arXiv.2405.13239},
	archivePrefix = {arXiv},
	eprint = {2405.13239},
	primaryClass = {q-bio.PE},
	adsurl = {https://ui.adsabs.harvard.edu/abs/2024arXiv240513239G}
}

@article{Kynaston2023,
  author  = {Kynaston, Joshua C. and Yates, Christian A. and Hekkink, Anna V. F. and Guiver, Chris},
  title   = {The regime-conversion method: a hybrid technique for simulating well-mixed chemical reaction networks},
  journal = {Frontiers in Applied Mathematics and Statistics},
  year    = {2023},
  volume  = {9},
  pages   = {1107441},
  doi     = {10.3389/fams.2023.1107441},
  url     = {https://www.frontiersin.org/articles/10.3389/fams.2023.1107441/full}
}

@article{Davis1984,
	author  = {Davis, Mark H. A.},
	title   = {Piecewise-Deterministic Markov Processes: A General Class of Non-Diffusion Stochastic Models},
	journal = {Journal of the Royal Statistical Society: Series B (Methodological)},
	year    = {1984},
	volume  = {46},
	number  = {3},
	pages   = {353--376},
	doi     = {10.1111/j.2517-6161.1984.tb01308.x}
}

@article{WinkelmannSchutte2017,
	author  = {Winkelmann, Stefanie and Sch\"utte, Christof},
	title   = {Hybrid models for chemical reaction networks: Multiscale theory and application to gene regulatory systems},
	journal = {The Journal of Chemical Physics},
	year    = {2017},
	volume  = {147},
	number  = {11},
	pages   = {114115},
	doi     = {10.1063/1.4986560}
}

@book{WinkelmannSchutte2020,
	author    = {Winkelmann, Stefanie and Sch\"utte, Christof},
	title     = {Stochastic Dynamics in Computational Biology},
	publisher = {Springer},
	series    = {Frontiers in Applied Dynamical Systems: Reviews and Tutorials},
	year      = {2020},
	volume    = {8},
	doi       = {10.1007/978-3-030-62387-6}
}

\newpage 
\appendix

\newcommand{\beginsupplement}{%
        \setcounter{table}{0}
        \renewcommand{\thetable}{S.I.\arabic{table}}%
        \setcounter{figure}{0}
        \renewcommand{\thefigure}{S.I.\arabic{figure}}%
        \setcounter{section}{0}
        \renewcommand{\thesection}{S.I.\arabic{section}}%
     }
\beginsupplement

\section{Backward equation for extinction probabilities} \label{sec:kolmogorov}
In this section, we derive the backwards Kolmogorov system given in Equation~\eqref{eq:bd-multitype-backward}.
Consider a time-inhomogeneous multitype Markov branching process on types $J$.
At time $s$, a single type-$i$ individual:
(i) dies at rate $\mu_i(s)$; 
(ii) gives birth to one additional individual of type $j$ at rate $\Lambda_{ij}(s)$.
Let $\mathbf{q}(s)=(q_i(s))_{i\in J}$ be the vector whose $i$th component is the
probability that a lineage started by a single type-$i$ individual at time $s$
is extinct by final time $T$, with terminal condition $q_i(T)=\mathds{1}\{i\notin J\}=0$.

By the backward Kolmogorov equation, conditioning on the first event in $(s,s+\mathrm{d}s)$,
\[
q_i(s)=\mu_i(s)\,\mathrm{d}s \cdot 1 \;+\; \sum_{j\in J}\Lambda_{ij}(s)\,\mathrm{d}s\; q_i(s+\mathrm{d}s)\,q_j(s+\mathrm{d}s)
\;+\; \Big(1-\mu_i(s)\mathrm{d}s-\!\sum_{j}\Lambda_{ij}(s)\mathrm{d}s\Big)\,q_i(s+\mathrm{d}s)
+ o(\mathrm{d}s).
\]
Subtracting $q_i(s+\mathrm{d}s)$, dividing by $\mathrm{d}s$ and letting $\mathrm{d}s\downarrow 0$ gives
\[
\frac{d}{ds}q_i(s) \;=\; \mu_i(s)\big(q_i(s)-1\big)
\;+\; q_i(s)\sum_{j}\Lambda_{ij}(s)\;-\; q_i(s)\sum_{j}\Lambda_{ij}(s)q_j(s).
\]
In vector form, using the Hadamard (element-wise) product $\circ$ and
$\mathbf{1}$ for the all-ones vector,
\[
\frac{d}{ds}\mathbf{q}(s)
= \boldsymbol{\mu}(s)\circ\big(\mathbf{q}(s)-\mathbf{1}\big)
+ \big(\Lambda(s)\mathbf{1}\big)\circ \mathbf{q}(s)
- \mathbf{q}(s)\circ\big(\Lambda(s)\mathbf{q}(s)\big),
\]
with terminal condition $\mathbf{q}(T)=\mathbf{0}$.
For a single type with rates $\lambda(s)$ and $\mu(s)$ this reduces to the Riccati ODE
$\;\dot q=\mu(s)\,(q-1)+\lambda(s)\,q\,(1-q)$.

\section{The Jump--Switch--Flow algorithm}\label{ap:jsf_algorithm}

This appendix provides a self-contained mathematical and algorithmic description of the Jump--Switch--Flow (JSF) method used throughout the main text. The presentation adapts the formulation of \citet{2024arXiv240513239G}; readers seeking further details, implementation notes, or open-source code are directed there.

\subsection{State space and partition}

Consider a reaction network with $n$ compartments $\mathbf V(t) = (V_1(t),\ldots,V_n(t))$ and $m$ reactions $\{\mathcal R_k\}_{k=1}^m$, each with propensity $\lambda_k(\mathbf V)$ and stoichiometric vector $\boldsymbol\eta_{\cdot k}\in\mathbb Z^n$. For each compartment $i$, JSF partitions its domain at a user-chosen threshold $\Omega_i\in\mathbb{Z}_{\geq 0}$ into a discrete part and a continuous part:
\begin{equation}
\mathcal V_{\Omega_i} \;=\; \{0,1,\ldots,\Omega_i\}\,\cup\,(\Omega_i,\infty).
\end{equation}
At any time $t$, the compartment is said to be \emph{jumping} if $V_i(t)\le\Omega_i$ and \emph{flowing} otherwise. Collecting indices, the state decomposes as $\mathbf V=(\mathbf V_J,\mathbf V_F)$, where $\mathbf V_J$ carries integer values and $\mathbf V_F$ real values. For notational simplicity we take a common threshold $\Omega$ below; the extension to species-specific $\Omega_i$ is immediate.

\subsection{Reaction partition}

Reactions are partitioned into a stochastic set $\mathcal S$ and a deterministic set $\mathcal S' = \mathcal R\setminus\mathcal S$:
\begin{equation}
\mathcal S \;=\; \bigl\{\mathcal R_k \;:\; \exists\, i \text{ with } V_i\in\mathbf V_J \text{ and } (\eta_{ik}\neq 0 \text{ or } \partial_{V_i}\lambda_k\neq 0)\bigr\}.
\end{equation}
A reaction is thus treated stochastically if it either changes a jumping compartment or has a propensity that depends on one. After reordering indices, the stoichiometric matrix admits the block form
\begin{equation}
\eta \;=\;
\left(\begin{array}{c|c}
\eta_{\mathcal S'} & \bar\eta_{\mathcal S} \\\hline
0 & \eta_{\mathcal S}
\end{array}\right),
\end{equation}
and between switching events the system obeys
\begin{align}
\frac{\mathrm d\mathbf V_F}{\mathrm d t} &= \eta_{\mathcal S'}\,\boldsymbol\lambda_{\mathcal S'}(\mathbf V) + \bar\eta_{\mathcal S}\,\boldsymbol\Lambda_{\mathcal S}(\mathbf V),\label{eq:jsf_flow_app}\\
\mathbf V_J(t) &= \mathbf V_J(t_0) + \eta_{\mathcal S}\int_{t_0}^{t} \boldsymbol\Lambda_{\mathcal S}(\mathbf V(s))\,\mathrm ds,\label{eq:jsf_jump_app}
\end{align}
where $\boldsymbol\Lambda_{\mathcal S}$ is a stochastic vector of Dirac spike trains indexed by the firing times of reactions in $\mathcal S$. There are three event types: \emph{jumps} (reactions in $\mathcal S$), \emph{flows} (continuous evolution of $\mathbf V_F$ under $\mathcal S'$), and \emph{switches} (threshold crossings that reclassify a compartment).

\subsection{Jump-clock construction}

Because $\mathbf V(t)$ varies continuously between jumps, the propensity $\lambda_k(\mathbf V(t))$ is time-dependent even in the absence of discrete events, and standard time-homogeneous Gillespie sampling is inapplicable. JSF samples the next firing time of each reaction $\mathcal R_k\in\mathcal S$ via a time-rescaling variant of the Next Reaction Method. The CDF of the next firing time, conditional on no firing since $t_0$, is
\begin{equation}
\mathrm{CDF}(t;k) \;=\; 1 - \exp\!\Bigl(-\!\int_{t_0}^{t}\lambda_k(\mathbf V(s))\,\mathrm ds\Bigr).
\end{equation}
Drawing $u_k\sim\mathrm{Unif}(0,1)$ and inverting gives the implicit equation $J_k(t_k)=0$ for the next firing time $t_k$, where
\begin{equation}
J_k(t) \;:=\; \log(u_k^{-1}) - \int_{t_0}^{t}\lambda_k(\mathbf V(s))\,\mathrm ds.
\label{eq:jump-clock}
\end{equation}
The quantity $J_k(t)$ is the \emph{jump clock} for reaction $\mathcal R_k$: initialised at $\log(u_k^{-1})>0$, it decays monotonically, and a firing is triggered when it reaches zero.

\subsection{Numerical integration of flows and clocks}

Between jumps, $\mathbf V_J$ is constant and $\mathbf V_F$ evolves under the ODE portion of Equation~\eqref{eq:jsf_flow_app}. JSF integrates this forward in time with a fixed step $\Delta t$ using an explicit scheme---forward Euler in the reference implementation \citep{2024arXiv240513239G}, higher-order schemes being a drop-in replacement:
\begin{equation}
\mathbf V_F(t+\Delta t) \;=\; \mathbf V_F(t) + \Delta t\,\eta_{\mathcal S'}\,\boldsymbol\lambda_{\mathcal S'}(\mathbf V(t)).
\end{equation}
Over the same step, each jump clock $J_k$ is decremented by the integral of $\lambda_k$. Expanding $\lambda_k(\mathbf V(t+s))$ to first order in $s$ yields
\begin{equation}
\Delta J_k \;=\; \int_0^{\Delta t}\lambda_k(\mathbf V(t+s))\,\mathrm ds \;\approx\; \frac{\Delta t}{2}\bigl(2\alpha + \beta\,\Delta t\bigr),
\end{equation}
with $\alpha = \lambda_k(\mathbf V(t))$ and $\beta = \boldsymbol\lambda_{\mathcal S'}(\mathbf V)^{\!\top}\eta_{\mathcal S'}^{\!\top}\nabla_{\mathbf V_F}\lambda_k$. If the provisional update $J_k(t)-\Delta J_k$ remains positive, no firing occurs during the step. If it becomes negative, the fractional time $\Delta\tau\in(0,\Delta t)$ at which the clock first reaches zero is obtained by solving the quadratic $2\Delta J_k = \Delta\tau\,(2\alpha + \beta\,\Delta\tau)$, giving
\begin{equation}
\Delta\tau \;=\;
\begin{cases}
\dfrac{\sqrt{\alpha^2 + 2\beta\,\Delta J_k}-\alpha}{\beta}, & \beta\neq 0,\\[4pt]
\dfrac{\Delta J_k}{\alpha}, & \beta = 0.
\end{cases}
\end{equation}
At $t+\Delta\tau$ the flow step is truncated, the stoichiometric update $\boldsymbol\eta_{\cdot k}$ is applied, and $J_k$ is reinitialised with a fresh $u_k\sim\mathrm{Unif}(0,1)$.

\subsection{Switching events}

Switching events reclassify a compartment between jumping and flowing. Two cases arise.

\paragraph{Jumping to flowing.} When a jump takes $V_i$ from $\Omega$ to $\Omega+1$, compartment $i$ is moved from $\mathbf V_J$ to $\mathbf V_F$. The continuous ODE for $V_i$ is initialised at $\Omega+1$ and integrated thereafter as part of $\mathbf V_F$; $\mathcal S$ and $\mathcal S'$ are updated accordingly.

\paragraph{Flowing to jumping.} When the flow takes $V_i$ below $\Omega$, compartment $i$ is moved from $\mathbf V_F$ to $\mathbf V_J$. In general the crossing value $\hat V_i\in(\Omega-1,\Omega]$ is non-integer, so JSF applies a probabilistic rounding that preserves the conditional mean:
\begin{equation}
V_i \;=\;
\begin{cases}
\lceil\hat V_i\rceil & \text{with probability } \hat V_i - \lfloor\hat V_i\rfloor,\\
\lfloor\hat V_i\rfloor & \text{with probability } 1 - (\hat V_i - \lfloor\hat V_i\rfloor).
\end{cases}
\end{equation}
This keeps $V_i\in\mathcal V_\Omega$ while ensuring $\mathbb E[V_i\mid\hat V_i]=\hat V_i$, preserving the first moment across the switch.

\subsection{Algorithm}

Algorithm~\ref{alg:jsf_pseudocode} summarises one simulation step. The algorithm mirrors the reference implementation in the accompanying open-source package \citep{2024arXiv240513239G}.

\begin{algorithm}[H]
\caption{One JSF step (advance from $t$ by at most $\Delta t$)}
\label{alg:jsf_pseudocode}
\begin{algorithmic}[1]
\State \textbf{Input:} state $\mathbf V$, partition $(J,F)$, jump clocks $\{J_k\}_{k\in\mathcal S}$, step $\Delta t$
\For{$k\in\mathcal S$} \Comment{Provisional clock updates}
    \State $\alpha_k \gets \lambda_k(\mathbf V)$; \quad $\beta_k \gets \boldsymbol\lambda_{\mathcal S'}(\mathbf V)^{\!\top}\eta_{\mathcal S'}^{\!\top}\nabla_{\mathbf V_F}\lambda_k$
    \State $\Delta J_k \gets \tfrac{\Delta t}{2}(2\alpha_k + \beta_k\Delta t)$
\EndFor
\State $k^\star\gets\arg\min_{k\in\mathcal S,\,J_k-\Delta J_k<0}\Delta\tau_k$ \Comment{Earliest firing, if any}
\If{no $k$ triggers (all $J_k-\Delta J_k\ge 0$)}
    \State $\mathbf V_F \gets \mathbf V_F + \Delta t\,\eta_{\mathcal S'}\boldsymbol\lambda_{\mathcal S'}(\mathbf V)$; \quad $J_k\gets J_k-\Delta J_k$ for all $k\in\mathcal S$
    \State $t\gets t+\Delta t$
\Else
    \State $\Delta\tau\gets\Delta\tau_{k^\star}$ \Comment{From quadratic root above}
    \State $\mathbf V_F \gets \mathbf V_F + \Delta\tau\,\eta_{\mathcal S'}\boldsymbol\lambda_{\mathcal S'}(\mathbf V)$ \Comment{Truncated flow step}
    \State Apply $\mathbf V\gets\mathbf V+\boldsymbol\eta_{\cdot k^\star}$; reinitialise $J_{k^\star}$ with fresh $u\sim\mathrm{Unif}(0,1)$
    \State $t\gets t+\Delta\tau$
\EndIf
\ForAll{$i$ with $V_i$ crossing $\Omega_i$} \Comment{Switching events}
    \State Update partition $(J,F)$; apply stochastic rounding if $F\!\to\! J$; update $\mathcal S,\mathcal S'$
\EndFor
\State \textbf{Output:} updated $(\mathbf V,J,F,\{J_k\},t)$
\end{algorithmic}
\end{algorithm}

\subsection{Relationship to the main-text formalism}

The piecewise-deterministic generator $\mathcal G_t$ in Equation~\eqref{eq:jsf_generator} is the infinitesimal description of the process implemented by Algorithm~\ref{alg:jsf_pseudocode}: the first summand encodes the jumps $\boldsymbol\Lambda_{\mathcal S}$ with propensities $\lambda_k$, and the gradient term encodes the flow $\eta_{\mathcal S'}\boldsymbol\lambda_{\mathcal S'}$. The partition $(J(t),F(t))$ in Equation~\eqref{eq:jsf_sets} is determined by the current compartment values relative to $\Omega$ and is updated at switching events as described above. The error analysis in Appendix~\ref{ap:error_bound} does not depend on the jump-clock construction or the numerical integration scheme; only on the fact that while $V_i\le\Omega$ the dynamics agree with the CTMC, and while $V_i>\Omega$ they agree with the ODE.

\section{Derivation of bound on extinction probability}\label{ap:error_bound}

In this section we provide a derivation of the bound on the extinction probability, $\Delta_{\Omega}(T)$, presented in the main text, Equation~\eqref{eq:gen-combined}. We first bound the ``late term'' error, arising from up-crossings after the critical time $s_*$ (see \ref{appendix:late_term}). We then provide a detailed derivation for the ``early term'' error, which requires bounding stochastic excursions in the death-dominated regime using martingale inequalities (see  \ref{appendix:early_term}). In \ref{appendix:both_terms} we combine these bounds and present the full algorithm. In \ref{appendix:heuristic_term} we present and and discuss the simplified heuristic bound, cf. Equation~\eqref{eq:practical-heuristic}.

\subsection{Late term bound: birth-dominated regime}\label{appendix:late_term}
Consider first the late term. Up-crossings occurring after $s_*$ represent the primary source of error because the birth-death balance favours persistence above the threshold. For the late term, we have:
\begin{equation}
	\Delta_{\Omega,\text{late}}(T) \leq \mathbb{E}\left[ [q_Y(\tau_\uparrow)]^{\Omega+1} \mathbf{1}\{s_* \leq \tau_\uparrow < T\} \right].
\end{equation}
While the process sits at $\Omega$, the total birth intensity is $\Omega \lambda (t)$ (there are $\Omega$ individuals, each giving birth at rate $\lambda(t)$). The jump event $\Omega \to \Omega +1$ is an inhomogeneous Poisson process, with density function
\begin{equation}
	w(t) = \Omega \lambda(t) \exp \left( -\Omega \int_0^t \lambda(u) du\right).
		\label{eq:up-crossing-density}
\end{equation}
At the up-crossing time $\tau_\uparrow = s$, the population is $Y(s) = \Omega + 1$, and by the branching property, the extinction probability is $[q_Y(s)]^{\Omega+1}$, giving:
\begin{equation}
	\Delta_{\Omega,\text{late}}(T) \leq \int_{s_*}^T w(s) [q_Y(s)]^{\Omega+1} \, ds.
	\label{eq:late-bound-integral}
\end{equation}

\subsection{Early term bound: death-dominated regime}\label{appendix:early_term}
For the early term, the error arises from the difference in return times of excursions between the ODE and stochastic models on the window $[0,s_*]$. In this window, assume a uniform negative drift margin:
\begin{equation}
	r(t) \leq -\rho < 0 \quad \text{for all } t\in[0,s_*], \qquad
	\rho := \inf_{t\in[0,s_*]} \bigl(-r(t)\bigr) > 0 .
	\label{eq:gen-rho-early}
\end{equation}
For an excursion that starts at time $s$ (i.e., $Y(s^-)=\Omega$ and $Y(s)=\Omega+1$), let $y^{(s)}$ denote the deterministic ODE trajectory solving $\dot y^{(s)}(t)=r(t)\,y^{(s)}(t)$ with $y^{(s)}(s)=\Omega+1$, and let $\sigma(s)$ be its (deterministic) return time to level $\Omega$, so $y^{(s)}(\sigma(s))=\Omega$. We define the deviation process on this excursion by
\begin{equation}
	M^{(s)}(t):=Y(t)-y^{(s)}(t),\qquad t\in[s,\sigma(s)].
\end{equation}
When $s$ is clear we omit the superscript and write $M(t)=Y(t)-y(t)$; expressions such as $M(\sigma(s))$ are understood with this convention. Let $\tau_{\uparrow,k}$ be the $k$-th up-crossing time $\Omega\!\to\!\Omega{+}1$ (with $\tau_{\uparrow,k}<s_*$), let $\tau_{\downarrow,k}$ be the corresponding \emph{stochastic} return time to $\Omega$, and let $\sigma(\tau_{\uparrow,k})$ be the \emph{deterministic} return time obtained by evolving the ODE segment from $y(\tau_{\uparrow,k})=\Omega{+}1$ until it hits $\Omega$:
\begin{equation}
	\int_{\tau_{\uparrow,k}}^{\sigma(\tau_{\uparrow,k})} r(u)\,du=\log\!\left(\frac{\Omega}{\Omega+1}\right).
\end{equation}
Write
\begin{equation}
	p(t):=\big[q_Y(t)\big]^{\Omega}
\end{equation}
for the extinction probability at horizon $T$ when starting from $\Omega$ individuals at time $t$; $p(t)$ is decreasing on $[0,s_*]$ and $L_{\mathrm{eff}}$-Lipschitz there, where
\begin{equation}
	L_{\mathrm{eff}} := \sup_{s\in[0,s_*]} |p'(s)|, 
	\qquad
	p'(s)=\Omega\,q_Y(s)^{\Omega-1}\left(\mu(s)\,(q_Y(s)-1)+\lambda(s)\,(q_Y(s)-q_Y(s)^2)\right).
	\label{eq:gen-Leff-early}
\end{equation}
The per-excursion contribution to the error is
\begin{equation}
	\theta(\tau_{\uparrow,k}) := \max\{0, p\big(\sigma(\tau_{\uparrow,k})\big)-p\big(\tau_{\downarrow,k}\big)\}
\end{equation}
(taking the positive part so ``helpful'' negative excursions do not cancel ``harmful'' ones). Summing over all early excursions,
\begin{equation}\label{eq:Delta-early-def}
	\Delta_{\Omega,\mathrm{early}}(T)
	=
	\mathbb{E}\left[\sum_{k=1}^{N_\uparrow}\theta(\tau_{\uparrow,k})\right]
	\leq \mathbb{E}[N_\uparrow]\;\sup_{s\in[0,s_*]} m(s)
\end{equation}
where $N_\uparrow$ is the number of up-crossings with start time $<s_*$ and 
\begin{equation}
	m(s)\;:=\;\mathbb{E}\!\left[\theta(s)\,\middle|\,\tau_{\uparrow,1}=s\right]
\end{equation}
is the expected per-excursion contribution when an excursion starts at time \(s\).

Let $N_\uparrow(t)$ count up-crossings $\Omega\!\to\!\Omega{+}1$ up to time $t$.
While $Y(t)=\Omega$, the instantaneous intensity of up-crossings is $\Omega\,\lambda(t)$,
so the compensator of $N_\uparrow$ on $[0,s_*]$ is $\int_0^{s_*}\Omega\,\lambda(t)\,\mathbf 1_{\{Y(t)=\Omega\}}\,dt$.
Taking expectations yields
\begin{equation}
	\mathbb{E}[N_\uparrow]=\mathbb{E}\!\left[\int_0^{s_*}\Omega\,\lambda(t)\,\mathbf 1_{\{Y(t)=\Omega\}}\,dt\right]
	\;\le\;\int_0^{s_*}\Omega\,\lambda(t)\,dt,
	\label{eq:gen-Enup}
\end{equation}
and, if desired, the crude simplification
$\mathbb{E}[N_\uparrow]\le \Omega\,\bar\lambda\,s_*$ with $\bar\lambda:=\sup_{t\in[0,s_*]}\lambda(t)$. To complete the bound, we must estimate $m(s)$, the expected per-excursion error contribution. The argument proceeds in three main steps: first, we use the smoothness of the extinction probability $p(t)$ to relate the error to the excursion's time duration; second, we bound this duration using a drift argument; and third, we bound the stochastic fluctuations using a martingale inequality.

Since $p(t)$ is $L_{\mathrm{eff}}$-Lipschitz and decreasing on $[0,s_*]$, we can bound the error contribution $\theta(s)$ by the difference in return times:
\begin{equation}
	\theta(s) \leq L_{\mathrm{eff}} \max\{0, \tau_{\downarrow}-\sigma(s)\}.
\end{equation}
The problem thus reduces to finding the expected time the stochastic process $Y(t)$ takes to return to $\Omega$ after the deterministic trajectory $y(t)$ has already done so at time $\sigma(s)$.

At $t=\sigma(s)$, the deterministic trajectory sits at level \(\Omega\), i.e. \(y(\sigma(s))=\Omega\). Any excess population in the stochastic process is given by $\phi(s):=\max\{0,\,Y(\sigma(s))-\Omega\}=\max\{0,\,M(\sigma(s))\}\le \sup_{t\in[s,\sigma(s)]}|M(t)|
$. For the process to return to $\Omega$, this excess must be depleted. During this return interval, $[\sigma(s), \tau_{\downarrow})$, the population size $Y(t)$ is at least $\Omega$ and the per-capita growth rate $r(t)$ is at most $-\rho$. Consequently, the total population drift $r(t)Y(t)$ is bounded above by $-\rho\Omega$ \footnote{Think of a tank that contains \(\phi(s)\) extra units and drains at a \emph{minimum} rate \(\rho\,\Omega\) whenever it is above \(\Omega\): at such a rate, the time to remove \(\phi(s)\) units cannot exceed \(\phi(s)/(\rho\,\Omega)\) on average.}. Applying the Optional Stopping Theorem to the process from time $\sigma(s)$ until the return time $\tau_{\downarrow}$ provides a bound on the expected time to drain this excess:
\begin{equation}
	\mathbb{E}\left[\tau_{\downarrow} - \sigma(s) \mid \mathcal{F}_{\sigma(s)}\right] \le \frac{\phi(s)}{\rho\Omega}.
\end{equation}
Taking expectations conditional on the excursion starting at $s$ and combining with the Lipschitz bound yields:
\begin{equation}
	m(s) = \mathbb{E}[\theta(s) \mid \tau_{\uparrow,1}=s] \le \frac{L_{\mathrm{eff}}}{\rho\Omega} \mathbb{E}[\phi(s) \mid \tau_{\uparrow,1}=s].
\end{equation}
The excess population $\phi(s)$ is itself bounded by the maximum deviation between the stochastic and deterministic paths, $M(t) := Y(t)-y(t)$, over the excursion interval. This leads to:
\begin{equation}
	m(s) \le \frac{L_{\mathrm{eff}}}{\rho\Omega} \mathbb{E}\left[\sup_{t \in [s, \sigma(s)]} \abs{M(t)} \middle| \tau_{\uparrow,1}=s\right].
	\label{eq:m_s_bound}
\end{equation}

The remainder of this subsection is dedicated to bounding the expected supremum $\mathbb{E}[\sup |M(t)|]$ using martingale analysis. The argument proceeds as follows:
\begin{itemize}
    \item First, we use a martingale decomposition and Gr\"onwall's inequality to relate the supremum of the full deviation, $\sup|M(t)|$, to the supremum of its core martingale component.
    \item Second, we apply the Burkholder-Davis-Gundy (BDG) inequality to bound the martingale supremum in terms of its predictable quadratic variation (PQV).
    \item Third, we derive a uniform, deterministic bound $V_\star$ for this PQV.
    \item Finally, we assemble these components to complete the early term bound.
\end{itemize}

\subsubsection{Martingale Decomposition and Gr\"onwall's Inequality}

Throughout this analysis, we analyse a single excursion that starts at time $s$ (an up-crossing $Y(s^-)=\Omega$, $Y(s)=\Omega{+}1$).
All martingales are taken with respect to the shifted filtration $(\mathcal F_{s+t})_{t\ge0}$, and all expectations may be read
conditionally on $\mathcal F_s$ (equivalently, on the event $\{\tau_{\uparrow,1}=s\}$ when we condition on the value of the first up-crossing time).
This conditioning does not change any inequality below: the Grönwall step is pathwise, and martingale inequalities such as BDG and Freedman hold with the same universal constants under the conditional law. \newline

Define the compensated c\`adl\`ag martingale increment on $[s,\infty)$:
\begin{equation}
	\tilde M^{(s)}(t) \;:=\; \tilde M(t)-\tilde M(s)
	\;=\; Y(t)-Y(s) - \int_s^t r(u)\,Y(u)\,du, \qquad t\ge s,
\end{equation}
so that $|\Delta\tilde M^{(s)}|\le 1$ and the predictable quadratic variation is
\begin{equation}
	\big\langle \tilde M^{(s)}\big\rangle_{[s,t]}
	\;=\; \int_s^t \big(\lambda(u)+\mu(u)\big)\,Y(u)\,du.
	\label{eq:quadratic_var}
\end{equation}
Let $y^{(s)}$ solve $\dot y^{(s)}(t)=r(t)\,y^{(s)}(t)$ with $y^{(s)}(s)=\Omega{+}1$, and set $M(t):=Y(t)-y^{(s)}(t)$.
Since excursions start at $Y(s)=y^{(s)}(s)=\Omega{+}1$, we have $M(s)=0$, and
\begin{equation}
	M(t)\;=\;\tilde M^{(s)}(t) \;+\; \int_s^t r(u)\,M(u)\,du.
\end{equation}
Define the envelopes $\phi(t):=\sup_{u\in[s,t]}|M(u)|$ and $\psi(t):=\sup_{u\in[s,t]}|\tilde M^{(s)}(u)|$. Then
\begin{equation}
	\phi(t)\;\le\;\psi(t) + \int_s^t |r(v)|\,\phi(v)\,dv,
\end{equation}
and by Gr\"onwall's inequality \citep{Gronwall1919,Walter1998}, for any $t\in[s,\sigma(s)]$,
\begin{equation}
	\sup_{u\in[s,t]} |M(u)|
	\;\le\; \exp\!\Big(\int_s^{t} |r(v)|\,dv\Big)\;
	\sup_{u\in[s,t]} |\tilde M^{(s)}(u)|.
\end{equation}
On the early window $r\le -\rho<0$ and using $\int_s^{\sigma(s)} r(u)\,du=\log\!\big(\Omega/(\Omega{+}1)\big)$, we obtain
\begin{equation}
	\sup_{u\le \sigma(s)} |M(u)|
	\;\le\; \frac{\Omega+1}{\Omega}\;
	\sup_{u\le \sigma(s)} |\tilde M^{(s)}(u)|.
\end{equation}
Taking \emph{conditional} expectations (given $\mathcal F_s$) gives
\begin{equation}\label{eq:expectation_of_M_cond}
	\E\!\left[\sup_{t\le \sigma(s)} |M(t)| \,\middle|\, \mathcal F_s\right]
	\;\le\; \frac{\Omega+1}{\Omega}\;
	\E\!\left[\sup_{t\le \sigma(s)}  |\tilde M^{(s)}(t)| \,\middle|\, \mathcal F_s\right].
\end{equation}

\subsubsection{Bounding the Martingale supremum via BDG}

Freedman's inequality yields for all $x>0$,
\begin{equation}
	\Pr \!\left(\sup_{t\le \sigma(s)} \abs{ \tilde M^{(s)}(t)} \ge x \,\middle|\, \mathcal F_s\right)
	\;\le\; 2 \exp \!\left( -\,\frac{x^2}{2\big(V + x/3\big)}\right),
\end{equation}
whenever $\langle \tilde M^{(s)}\rangle_{[s,\sigma(s)]}\le V$ almost surely (bounded jumps $\le 1$).
Consequently,
\begin{equation}
	\E \!\left[\sup_{t\le \sigma(s)} \abs{\tilde M^{(s)}(t)} \,\middle|\, \mathcal F_s\right]
	\;\le\; \int_0^{\infty} 2 \exp \!\left( -\frac{x^2}{2(V + x/3)}\right) dx .
\end{equation}

Rather than handling the long tail numerically, one can conveniently bypass it with the Burkholder–Davis–Gundy (BDG) inequality \citep{Burkholder1966,Davis1970,BurkholderGundy1970} . Taking the 1st moment ($p=1$) yields a universal constant $C_{\mathrm{BDG}}$ (we take $C_{\mathrm{BDG}}=2$) such that 
\begin{equation}\label{eq:bdg_martingale_cond}
	\mathbb{E} \left[\sup_{t\le \sigma(s)} \left|\tilde{M}^{(s)}(t)\right| \,\bigg|\, \mathcal{F}_s\right]
	\;\le\; C_{\mathrm{BDG}}\ \mathbb{E}\left[\langle \tilde{M}^{(s)}\rangle_{[s,\sigma(s)]}^{1/2} \,\bigg| \, \mathcal{F}_s\right].
\end{equation}
If, in addition, we have the uniform bound (see Appendix \ref{ap:martingale_variance})
\begin{equation}
	\mathbb{E}\left[\langle \tilde{M}^{(s)}\rangle_{[s,\sigma(s)]} \,\bigg| \, \mathcal{F}_s\right] \;\le\; V_\star \qquad \text{for all } s\in[0,s_*],
\end{equation}

then Jensen's inequality (concavity of $\sqrt{\cdot}$) yields
\begin{equation}\label{eq:bdg_martingale_uniform}
	\E \!\left[\sup_{t\le \sigma(s)} \abs{\tilde M^{(s)}(t)} \,\middle|\, \mathcal F_s\right]
	\;\le\; C_{\mathrm{BDG}}\ \sqrt{V_\star}.
\end{equation}

Combining Equations~\eqref{eq:expectation_of_M_cond} and \eqref{eq:bdg_martingale_uniform},
\begin{equation}
	\E\!\left[\sup_{t\le \sigma(s)} |M(t)| \,\middle|\, \mathcal F_s\right]
	\;\le\; \frac{\Omega+1}{\Omega}\, C_{\mathrm{BDG}}\ \sqrt{V_\star}.
	\label{eq:bdg_final_uniform}
\end{equation}
uniformly in $s\in[0,s_*]$. This is the estimate invoked in the early-term bound in the main text.

\subsubsection{Bounding the predictable quadratic variation} \label{ap:martingale_variance}

This section is concerned with calculating $V_\star$, a deterministic bound on the conditional expected predictable quadratic variation (PQV) accumulated over any early excursion. $V_\star$ plays a crucial role in controlling the early excursion behaviour of the birth--death process. The calculation of $V_\star$ allows us to establish concrete bounds on the martingale deviation through the application of the BDG inequality, cf. Equation~\eqref{eq:gen-early-bound-final-cleaned}

In the general time-inhomogeneous birth--death case, on an excursion started at time $s$ with $Y(s)=\Omega+1$, the compensated jump martingale
$
\tilde M^{(s)}(t) := Y(t)-Y(s)-\int_s^t r(u)Y(u)\,du
$
has PQV
\begin{equation}
	\langle \tilde M^{(s)}\rangle_{[s,t]}
	= \int_s^t \big(\lambda(u)+\mu(u)\big)\,Y(u)\,du,
	\qquad r(u)=\lambda(u)-\mu(u).
\end{equation}
For $t\in[s,\sigma(s)]$,
\begin{equation}
	\mathbb E\!\left[\langle \tilde M^{(s)}\rangle_{[s,\sigma(s)]}\,\middle|\,\mathcal F_s\right]
	= \int_s^{\sigma(s)} \big(\lambda(u)+\mu(u)\big)\,\mathbb E\!\left[Y(u)\,\middle|\,\mathcal F_s\right] du.
\end{equation}
Since the per-capita rates depend only on time, $\mathbb E[Y(u)\mid\mathcal F_s]$ solves the same linear ODE as the deterministic envelope $y^{(s)}$:
$
\frac{d}{du}\,y^{(s)}(u)=r(u)\,y^{(s)}(u),\ y^{(s)}(s)=\Omega+1,
$
hence
\begin{equation}
	\mathbb E\!\left[\langle \tilde M^{(s)}\rangle_{[s,\sigma(s)]}\,\middle|\,\mathcal F_s\right]
	= \int_s^{\sigma(s)} \big(\lambda(u)+\mu(u)\big)\,y^{(s)}(u)\,du
	\;\le\; V_\star,
	\label{eq:Vstar-general}
\end{equation}
with
\begin{equation}
	V_\star \;:=\; \sup_{s\in[0,s_*]}\int_s^{\sigma(s)} \big(\lambda(u)+\mu(u)\big)\,y^{(s)}(u)\,du.
	\label{eq:Vstar-def}
\end{equation}
Using the early-window negativity $r\le -\rho<0$ and $y^{(s)}(u)\le (\Omega+1)e^{-\rho(u-s)}$, we get the convenient crude bound
\begin{equation}
	V_\star
	\;\le\; \sup_{t\in[0,s_*]}\big(\lambda(t)+\mu(t)\big)\,(\Omega+1)\!\int_0^\infty e^{-\rho w}\,dw
	\;=\; \frac{\Omega+1}{\rho}\ \sup_{t\in[0,s_*]}\big(\lambda(t)+\mu(t)\big).
	\label{eq:gen-Vstar}
\end{equation}

\subsubsection{Completing the early term bound}

We now have all the pieces to bound the per-excursion error $m(s)$. Substituting the martingale bound from Equation~\eqref{eq:bdg_final_uniform} into our expression for $m(s)$ in Equation~\eqref{eq:m_s_bound} yields a uniform bound on the per-excursion error:
\begin{equation}
	\sup_{s\in[0,s_*]} m(s) \le \frac{L_{\mathrm{eff}}}{\rho\Omega} \left( \frac{\Omega+1}{\Omega}\, C_{\mathrm{BDG}}\ \sqrt{V_\star} \right).
\end{equation}
Finally, substituting this bound for $m(s)$ into Equation~\eqref{eq:Delta-early-def} along with the estimate for $\mathbb{E}[N_\uparrow]$ from Equation~\eqref{eq:gen-Enup} yields the final estimate for the total early-term error:
\begin{equation}
	\Delta_{\Omega,\mathrm{early}}(T) \leq \left(\int_0^{s_*}\Omega\,\lambda(t)\,dt\right) \frac{L_{\mathrm{eff}}}{\rho} \frac{\Omega+1}{\Omega} C_{\mathrm{BDG}} \sqrt{V_\star}. \label{eq:gen-early-bound-final-cleaned}
\end{equation}

\subsection{Final combined bound and algorithm}\label{appendix:both_terms}
Putting together Equations~\eqref{eq:gen-early-bound-final-cleaned} and \eqref{eq:late-bound-integral} yields
\begin{equation}
	\Delta_\Omega(T) \le \left(\int_0^{s_*}\Omega\lambda(t)\,dt\right) \frac{L_{\text{eff}}}{\rho} \frac{\Omega+1}{\Omega} C_{\text{BDG}} \sqrt{V_\star} + \int_{s_*}^T \Omega\lambda(s)e^{-\Omega\int_0^s\lambda}[q(s)]^{\Omega+1}\,ds. \label{eq:gen-combined2}
\end{equation}
Algorithm \ref{alg:full-bound} outlines the complete procedure for computing the formal bound.
\begin{algorithm}[H]
	\caption{Full Error Bound for Birth-Death Systems}
	\label{alg:full-bound}
	\begin{algorithmic}[1]
		\State \textbf{Input:} Threshold $\Omega$, time horizon $T$, birth rate $\lambda(t)$, death rate $\mu(t)$
		
		\Comment{\textit{Part 1: Preliminary Calculations}}
		\State Compute net growth rate: $r(t) = \lambda(t) - \mu(t)$
		\State Find critical time: $t_c = \inf\{t \geq 0 : r(t) \geq 0\}$
		\State Solve for point of no return $s_*$ from: $\int_{s_*}^{t_c} r(u) \, du = \log\left(\frac{\Omega}{\Omega+1}\right)$
		\State Solve Riccati equation backward for extinction probability $q_Y(s)$ on $[0, T]$: 
		\Statex \hspace{\algorithmicindent} $\frac{dq_Y}{ds} = \mu(s)(q_Y - 1) + \lambda(s)(q_Y - q_Y^2)$, with terminal condition $q_Y(T) = 0$
		
		\vspace{1mm}
		\Comment{\textit{Part 2: Early Excursion Error Term on $[0, s_*]$}}
		\State Compute negative drift margin: $\rho = \inf_{t\in[0,s_*]} (-r(t))$
		\State Compute expected number of up-crossings: $\mathbb{E}[N_\uparrow] = \int_0^{s_*} \Omega \lambda(t) \, dt$
		\State Define extinction probability from $\Omega$: $p(t) = [q_Y(t)]^\Omega$
		\State Compute effective Lipschitz constant: $L_{\text{eff}} = \sup_{s\in[0,s_*]} |p'(s)|$
		\State Compute an upper bound $V_\star$ for the martingale quadratic variation on $[0, s_*]$
		\State Assemble early term bound (where $C_{\text{BDG}}$ is the Burkholder-Davis-Gundy constant):
		\Statex \hspace{\algorithmicindent} $\Delta_{\Omega,\text{early}}(T) = \mathbb{E}[N_\uparrow] \cdot \frac{L_{\text{eff}}}{\rho} \frac{\Omega+1}{\Omega} C_{\text{BDG}} \sqrt{V_\star}$
		
		\vspace{1mm}
		\Comment{\textit{Part 3: Late Excursion Error Term on $[s_*, T]$}}
		\State Compute the up-crossing time density: $w(s) = \Omega \lambda(s) \exp\left(-\Omega \int_0^s \lambda(u) \, du\right)$
		\State Compute the late term bound via integration: 
		\Statex \hspace{\algorithmicindent} $\Delta_{\Omega,\text{late}}(T) = \int_{s_*}^T w(s) [q_Y(s)]^{\Omega+1} \, ds$
		
		\vspace{1mm}
		\Comment{\textit{Part 4: Final Bound}}
		\State Combine the two terms: $\Delta_\Omega(T) = \Delta_{\Omega,\text{early}}(T) + \Delta_{\Omega,\text{late}}(T)$
		\State \textbf{Output:} Full error bound $\Delta_\Omega(T)$
	\end{algorithmic}
\end{algorithm}

\subsection{Heuristic late-only bound }\label{appendix:heuristic_term}

As a fast and implementable approximation, we retain only the \emph{late} contribution (up-crossings after $s_*$) and neglect the early term. This is often appropriate in regimes where negative drift on $[0,s_*]$ quickly corrects early excursions, so most discrepancy arises when a path up-crosses and fails to return before the drift turns nonnegative.

The key practical insight is that in most birth-death systems, the late term dominates the error budget. When the death rate significantly exceeds the birth rate during the early period ($r(t) \ll 0$ for $t \in [0, s_*]$), excursions above the threshold are rapidly corrected, making the early term contribution negligible.

Starting from the general late bound with the nonhomogeneous up-crossing density $w(s)=\Omega\,\lambda(s)\exp\{-\Omega\int_0^s\lambda(u)\,du\}$, we have:
\begin{equation}
	\Delta_{\Omega,\text{late}}(T) \leq \int_{s_*}^T \Omega\,\lambda(s)\,e^{-\Omega\int_0^s\lambda(u)\,du}\,[q_Y(s)]^{\Omega+1}\,ds.
	\label{eq:late-heuristic-integral}
\end{equation}

We can obtain a simple analytic upper bound by replacing $\lambda$ with $\bar\lambda:=\sup_{[0,T]}\lambda$ and using that $q_Y(\cdot)$ is decreasing:
\begin{equation}
	\Delta_{\Omega,\text{late}}(T) \leq \int_{s_*}^T \Omega\,\bar{\lambda}\,e^{-\Omega\bar{\lambda} s}\,[q_Y(s)]^{\Omega+1}\,ds
	\leq [q_Y(s_*)]^{\Omega+1}.
	\label{eq:late-analytic-bound}
\end{equation}

If a more conservative bound is preferred, we can replace the exponent $\Omega + 1$ with $\Omega$:
\begin{equation}
	\Delta_\Omega(T) \lesssim [q_Y(s_*)]^\Omega.
	\label{eq:conservative-heuristic}
\end{equation}
This bound is looser but may be more convenient for reporting single-number accuracy guarantees.

\section{Calculating $s_*$ (Lambert-$W$)}\label{ap:getting_s_star}

We want to find the value $s_* \in [0, t_c]$ that satisfies the integral criterion:
\begin{equation}
	\int_{s_*}^{t_c} r(u)\,du = \log\left(\frac{\Omega}{\Omega+1}\right)
\end{equation}
where the rate function is $r(t) = \alpha - \beta x_2(0)e^{-\gamma t}$. The upper limit of integration, $t_c$, is the time at which $r(t_c)=0$, given by $t_c = \frac{1}{\gamma}\log(\frac{B}{\alpha})$, assuming $B := \beta x_2(0) > \alpha$.

\subsection{Numerical solution}
We can solve for $s_*$ by finding the root of a one-dimensional, monotone equation. Let's define a function $F(s)$ on the interval $[0, t_c]$ such that $s_*$ is its root. First, let $L := \log(\frac{\Omega}{\Omega+1})$, which is a negative constant. The integral can be calculated explicitly:
\begin{equation}
	I(s) := \int_s^{t_c} r(u)\,du = \left[\alpha u + \frac{B}{\gamma}e^{-\gamma u}\right]_s^{t_c} = \alpha(t_c-s) + \frac{B}{\gamma}\left(e^{-\gamma t_c} - e^{-\gamma s}\right)
\end{equation}
Using the fact that $e^{-\gamma t_c} = \alpha/B$, this simplifies to:
\begin{equation}
	I(s) = \alpha(t_c-s) + \frac{\alpha}{\gamma} - \frac{B}{\gamma}e^{-\gamma s}
\end{equation}
The value $s_*$ is the solution to $I(s_*) = L$. We can define the function to solve as:
\begin{equation}
	F(s) := I(s) - L = 0
\end{equation}
This function is continuous and strictly increasing on $[0, t_c]$ because its derivative is positive for $s < t_c$:
\begin{equation}
	F'(s) = -r(s) = -(\alpha - Be^{-\gamma s}) = Be^{-\gamma s} - \alpha > 0
\end{equation}
The values at the interval's endpoints are $F(t_c) = I(t_c) - L = 0 - L = -L > 0$ and $F(0) = \alpha t_c + \frac{\alpha}{\gamma} - \frac{B}{\gamma} - L$. Since $F(s)$ is monotone and $F(t_c)>0$, a unique root $s_* \in [0, t_c)$ exists if and only if $F(0) \le 0$. If $F(0) > 0$, no such root exists, and we take $s_* := 0$. Given a bracket $[a, b] = [0, t_c]$ where $F(a) \le 0 < F(b)$, the unique root can be found efficiently using (e.g.) bisection method.

\subsection{Closed-form solution}
Alternatively, the equation can be solved analytically to yield a closed-form expression for $s_*$. Starting from the integrated equation $I(s_*)=L$:
\begin{equation}
	\alpha(t_c-s_*) + \frac{\alpha}{\gamma} - \frac{B}{\gamma}e^{-\gamma s_*} = L
\end{equation}
We rearrange and substitute $t_c = \frac{1}{\gamma}\log(\frac{B}{\alpha})$:
\begin{equation}
	\frac{\alpha}{\gamma}\log\left(\frac{B}{\alpha}\right) + \frac{\alpha}{\gamma} - L = \alpha s_* + \frac{B}{\gamma}e^{-\gamma s_*}
\end{equation}
Multiplying by $\frac{\gamma}{\alpha}$ gives:
\begin{equation}
	\log\left(\frac{B}{\alpha}\right) + 1 - \frac{\gamma L}{\alpha} = \gamma s_* + \frac{B}{\alpha}e^{-\gamma s_*}
\end{equation}
Let's introduce a change of variables with $u := \frac{B}{\alpha}e^{-\gamma s_*}$. This implies $\gamma s_* = \log(\frac{B}{\alpha u}) = \log(\frac{B}{\alpha}) - \log(u)$. Substituting these into the equation yields:
\begin{equation}
	\log\left(\frac{B}{\alpha}\right) + 1 - \frac{\gamma L}{\alpha} = \left(\log\left(\frac{B}{\alpha}\right) - \log u\right) + u
\end{equation}
The $\log(\frac{B}{\alpha})$ terms cancel, leaving the transcendental equation:
\begin{equation}
	u - \log u = C, \qquad \text{where} \qquad C := 1 - \frac{\gamma L}{\alpha}
\end{equation}
This equation can be solved using the Lambert-W function. By rearranging, we get $ue^{-u} = e^{-C}$. The solution for $u$ is $u = -W(-e^{-C})$. Since $L<0$, we have $C > 1$, which means there are two possible real solutions, corresponding to the $W_0$ and $W_{-1}$ branches of the function. The physically relevant solution must satisfy $s_* \in [0, t_c)$, which corresponds to $u \in [1, B/\alpha]$. This requires the $W_{-1}$ branch. \newline 

Solving for $s_*$ from $u = \frac{B}{\alpha}e^{-\gamma s_*}$ gives the final closed-form expression:
\begin{equation}\label{eq:sstar-Lambert}
	s_*
	= -\frac{1}{\gamma}\log\left(\frac{\alpha}{B}\left[-W_{-1}\left(-e^{-C}\right)\right]\right)
	\in [0, t_c)
\end{equation}


\section{Lotka--Volterra model: Detailed derivations}\label{ap:lotka_volt}

This appendix provides explicit formulas for the Lotka--Volterra predator--prey system introduced in Section~\ref{sec:lv-case-study}. Recall the rates from Equation~\eqref{eq:lv-rates}:
\begin{equation}
\dot{x}_2(t)=-\gamma x_2(t),\qquad x_2(t)=x_{2,0}e^{-\gamma t},\qquad
\lambda(t)\equiv\alpha,\quad \mu(t)=\beta x_2(t),\quad r(t)=\alpha-\beta x_2(t),
\tag{\ref{eq:lv-rates}}
\end{equation}
where $\alpha,\beta,\gamma>0$ are fixed parameters, the prey $Y=X_1$ is the focal species, and the predator $X_2$ evolves deterministically.

\subsection{Derivation of the LV bound}

The sign change of $r(t)=\alpha-\beta x_{2,0}e^{-\gamma t}$ determines the critical time $t_c$ from $r(t_c)=0$:
\begin{equation}
t_c=\frac{1}{\gamma}\log\!\left(\frac{\beta x_{2,0}}{\alpha}\right)\quad\text{if } \beta x_{2,0}>\alpha,\quad\text{else }t_c=0.
\label{eq:lv-tc}
\end{equation}
The point of no return $s_\star\in[0,t_c)$ is defined by
\begin{equation}
\int_{s_\star}^{t_c} \big(\alpha-\beta x_{2,0}e^{-\gamma u}\big)\,du
=\log\!\left(\frac{\Omega}{\Omega+1}\right),
\end{equation}
with a closed form via the Lambert $W$ function given in Appendix~\ref{ap:getting_s_star}.

With $\lambda\equiv\alpha$ and $\mu=\beta x_2$, the backward Kolmogorov equation for the one-lineage extinction probability is
\begin{equation}
\frac{dq_Y}{ds}=\mu(s)(q_Y-1)+\lambda(s)(q_Y-q_Y^2)
=\beta x_{2,0}e^{-\gamma s}(q_Y-1)+\alpha (q_Y-q_Y^2),\quad q_Y(T)=0.
\label{eq:lv-riccati}
\end{equation}
For initial condition $Y(0)=\Omega$, the exact extinction probability by horizon $T$ is $P_{\mathrm{ext}}(T)=[q_Y(0)]^\Omega$.

The up-crossing intensity on $[s_\star,T]$ yields
\begin{equation}
\Delta_{\Omega,\mathrm{late}}(T) \le \int_{s_\star}^{T} \Omega\,\lambda(s)\,[q_Y(s)]^{\Omega+1}\,ds
 = \int_{s_\star}^{T} \Omega\,\alpha\,[q_Y(s)]^{\Omega+1}\,ds.
\label{eq:lv-late}
\end{equation}
A one-number summary follows from the monotonicity of $q_Y$:
\begin{equation}
\Delta_{\Omega,\mathrm{late}}(T)
 \le [q_Y(s_\star)]^{\Omega+1}\,\min\{1, \Omega\alpha(T-s_\star)\}
 \le [q_Y(s_\star)]^{\Omega+1}.
\label{eq:lv-heuristic}
\end{equation}

On $[0,s_\star]$, the growth rate remains strictly negative. Define the uniform negative margin
\begin{equation}
\rho:=\inf_{t\in[0,s_\star]}(\mu(t)-\lambda(t))
=\inf_{t\in[0,s_\star]}(\beta x_{2,0}e^{-\gamma t}-\alpha)
=\beta x_{2,0}e^{-\gamma s_\star}-\alpha>0.
\end{equation}
Then $\mathbb{E}[N_\uparrow]\le \int_0^{s_\star}\Omega\lambda(t)\,dt=\Omega\alpha s_\star$, and the LV version of the early bound is
\begin{equation}
\Delta_{\Omega,\mathrm{early}}(T)
 \le \underbrace{\Omega\,\alpha\,s_\star}_{\mathbb{E}[N_\uparrow]}\,
\frac{L_{\mathrm{eff}}}{\rho}\,\frac{\Omega+1}{\Omega}\,C_{\mathrm{BDG}}\,\sqrt{V_\star},
\label{eq:lv-early}
\end{equation}
with a convenient deterministic bound (see Appendix~\ref{ap:variance_LV})
\begin{equation}
V_\star \le \frac{\Omega+1}{\rho} \sup_{t\in[0,s_\star]}\big(\lambda(t)+\mu(t)\big)
=\frac{\Omega+1}{\rho} \sup_{t\in[0,s_\star]}\big(\alpha+\beta x_{2,0}e^{-\gamma t}\big)
=\frac{\Omega+1}{\rho} \big(\alpha+\beta x_{2,0}\big).
\label{eq:lv-Vstar}
\end{equation}

Adding Equations~\eqref{eq:lv-late} and \eqref{eq:lv-early} gives the Lotka--Volterra instance of the general bound:
\begin{equation}
\Delta_\Omega(T)
 \le \Big(\Omega\alpha s_\star\Big)\,\frac{L_{\mathrm{eff}}}{\rho}\,\frac{\Omega+1}{\Omega}\,C_{\mathrm{BDG}}\sqrt{V_\star}
+\int_{s_\star}^{T}\Omega\alpha\,[q_Y(s)]^{\Omega+1}\,ds.
\end{equation}

\subsection{Variance term $V_\star$: exact and approximate formulas}\label{ap:variance_LV}

For completeness, we derive the predictable quadratic variation bound $V_\star$ specialised to the deterministic predator case. With $r(t)=\alpha-\beta x_2(t)$ and $x_2(t)=x_{2,0}e^{-\gamma t}$, the exact expression is
\begin{equation}
V_\star=\sup_{s\in[0,s_\star]}\int_s^{\sigma(s)}\!\big(\alpha+\beta x_2(u)\big)\,y^{(s)}(u)\,du.
\label{eq:Vstar-PP-exact}
\end{equation}
There are two useful forms:

\begin{enumerate}
	\item \textbf{Option A (simple bound).} \newline 
	Using $y^{(s)}(u)\le (\Omega+1)e^{-\rho(u-s)}$ and $\sup_{[0,s_\star]}(\alpha+\beta x_2)<\infty$,
	\begin{equation}
		V_\star \le \frac{\Omega+1}{\rho} \sup_{t\in[0,s_\star]}\big(\alpha+\beta x_2(t)\big).
		\label{eq:Vstar-PP-simple}
	\end{equation}
	\item \textbf{Option B (tighter expression).} Let
	\begin{equation}
		R(t)=\int_0^t r(u)\,du,\quad
		G(t)=(\alpha+\beta x_2(t))e^{R(t)},\quad
		H(t)=\int_0^t G(u)\,du.
	\end{equation}
	Then $y^{(s)}(u)=(\Omega+1)e^{R(u)-R(s)}$ and $R(\sigma(s))-R(s)=\log\!\big(\Omega/(\Omega+1)\big)$, so
	\begin{equation}
		\int_s^{\sigma(s)}\!\big(\alpha+\beta x_2(u)\big)\,y^{(s)}(u)\,du
		=(\Omega+1)e^{-R(s)}\big(H(\sigma(s))-H(s)\big),
	\end{equation}
	hence
	\begin{equation}
		V_\star=\sup_{s\in[0,s_\star]} (\Omega+1)\,e^{-R(s)}\big(H(\sigma(s))-H(s)\big).
		\label{eq:Vstar-PP-tight}
	\end{equation}
	
\end{enumerate}

\end{document}